\documentclass[12pt]{article}

\usepackage{graphics}
\usepackage{amssymb}

\renewcommand{\theequation}{\arabic{section}.\arabic{equation}}

\newcommand{\e}{\mathrm{e}}
\newcommand{\D}{\mathrm{d}}
\newcommand{\C}{\mathbb{C}}

\newcommand{\R}{\mathbb{R}}
\newtheorem{claim}{Claim}[section]
\newtheorem{theorem}[claim]{Theorem}
\newtheorem{lemma}[claim]{Lemma}
\newtheorem{remark}[claim]{Remark}
\newtheorem{remarks}[claim]{Remarks}
\newtheorem{proposition}[claim]{Proposition}
\newenvironment{proof}[1][Proof]{\textsl{#1.} }{\ \rule{0.5em}{0.5em}}

\begin{document}

\title{Schr\"odinger operators with singular interactions:
a model of tunneling resonances}
\author{P.~Exner$^{a,b}$ and S.~Kondej$^c$}
\date{}
\maketitle

\begin{quote}
{\small \em a) Nuclear Physics Institute, Academy of Sciences,
25068 \v Re\v z \\ \phantom{a) }near Prague, Czech Republic
\\
b) Doppler Institute, Czech Technical University,
B\v{r}ehov{\'a}~7, \\ \phantom{a) }11519 Prague, Czech Republic
\\
c) Institute of Physics, University of Zielona G\'{o}ra, ul.
Szafrana 4a, \\ \phantom{a) } 65246 Zielona G\'{o}ra, Poland
\\
\phantom{a) }\texttt{exner@ujf.cas.cz},
\texttt{skondej@if.uz.zgora.pl} }
\end{quote}

\begin{quote}
{\small We discuss a generalized Schr\"odinger operator in
$L^2(\R^d),\, d=2,3$, with an attractive singular interaction
supported by a $(d-1)$-dimensio\-nal hyperplane and a finite
family of points. It can be regarded as a model of a leaky quantum
wire and a family of quantum dots if $d=2$, or surface waves in
presence of a finite number of impurities if $d=3$. We analyze the
discrete spectrum, and furthermore, we show that the resonance
problem in this setting can be explicitly solved; by
Birman-Schwinger method it is cast into a form similar to the
Friedrichs model.}
\end{quote}


\section{Introduction}
\label{introd} \setcounter{equation}{0}

The subject of this paper is a nonrelativistic quantum Hamiltonian
in $L^2(\R^d)$, $d=2,3$, with a singular interaction supported by
a set consisting of two parts. One is a flat manifold of dimension
$d\!-\!1$, i.e. a line for $d=2$ and a plane for $d=3$, the other
is a finite family of points situated in general in the complement
to the manifold. The corresponding generalized Schr\"odinger
operator can be formally written as
 \begin{equation} \label{forHam}
 -\Delta -\alpha \delta (x-\Sigma ) +\sum_{i=1}^n \tilde\beta_i
 \delta(x-y^{(i)})\,,
 \end{equation}
where $\alpha>0$, $\,\Sigma :=\{(x_1,0);\,x_1\in\R^{d-1}\}$, and
$y^{(i)}\in \R^d \setminus \Sigma$; the formal coupling constants
of the $d$-dimensional $\delta$ potentials are marked by tildas
because they are not the proper parameters to be used; we will
discuss this point in more detail below.

The first question to be posed is about a physical significance of
such a Hamiltonian. Operators of the type (\ref{forHam}) or
similar have been studied recently with the aim to describe
nanostructures which are ``leaky'' in the sense that they do not
neglect quantum tunneling -- cf.~\cite{Ex2, EI, EK, EN, ET, EY1,
EY2} and references therein. In this sense we can regard the
present model with $d=2$ as an idealized description of a quantum
wire and a collection of quantum dots which are spatially
separated but close enough to each other so that electrons are
able to pass through the classically forbidden zone separating
them. Similarly the three-dimensional case can be given
interpretation as a description of surface states under influence
of a finite number of point perturbations.

We will ask first about the discrete spectrum of the Hamiltonian
(\ref{forHam}). It will be demonstrated to be always nonempty and
properties of the eigenvalues in terms of the model parameters
will be derived, which complements the existing knowledge about
the discrete spectrum of such generalized Schr\"odinger operators
derived in the mentioned papers and earlier, e.g., in \cite{BT}.

Our main concern in this paper, however, is the scattering within
our model, in particular, the question about existence of the
resonances. It is obvious that this is an important problem for
generalized Schr\"odinger operators with the interaction supported
by a non-compact manifold of a lower dimension, of which a little
is known at present. The simple form of the interaction support,
$\Sigma \cup\Pi$ with $\Pi:= \{y^{(i)}\}$, will allow us to
analyze the scattering for the operator (\ref{forHam}). We will
achieve that by using the generalized Birman-Schwinger method
which makes it possible to convert the original PDE problem into a
simpler equation which in the present situation is in part
integral, in part algebraic. The main insight is that the method
works not only for the discrete spectrum but it can be used also
to find singularities of the analytically continued resolvent. The
problem can be then reduced to a finite rank perturbation of
eigenvalues embedded in the continuous spectrum, i.e. something
which calls to mind the celebrated Friedrichs model --
cf.~\cite{Fr} or \cite[Sec.~3.2]{Ex1}.

We will pay most attention to the two-dimensional case. In the
next section we will first explain how the operator (\ref{forHam})
should be properly defined, then we derive a Birman-Schwinger-type
expression for its resolvent. Using this information we discuss in
Sec.~3 the discrete spectrum, first for $n=1$, then for a pair of
point perturbations showing how embedded eigenvalues due to
symmetry may arise, and finally for a general $n$. In Sec.~4 we
tackle the resonance problem using the mentioned analytical
continuation of the resolvent. For simplicity we consider only the
cases of a single perturbation, where the resonance width is found
to be exponentially in terms of the distance between the line and
the point, and of a pair of them to illustrate how resonances can
arise from symmetry breaking. We will also treat the same problem
with $n=1$ from other point of views: as a scattering of a
particle transported along the line and as a decaying unstable
system. Finally in Sec.~5 we investigate the three-dimensional
case. Since the analysis is similar, we restrict ourselves to
describing the features which are different for $d=3$.


\section{The Hamiltonian for $d=2$}
\label{line+point} \setcounter{equation}{0}

\subsection{Definition of Hamiltonian}\label{defHam}

If $d=2$ the interaction is supported by $\Sigma \cup\Pi$ with
$\Sigma :=\{(x_{1},0);\,x_{1}\in\R\}$ and $\Pi:=
\{y^{(i)}\}_{i=1}^n\,$, where $y^{(i)}\in \R^2 \setminus \Sigma $.
For simplicity we also put $L^{2}\equiv L^{2}(\R^{2})$. The most
natural way to find a self-adjoint realization of the formal
expression (\ref{forHam}) is to construct the Laplace operator
with appropriated boundary conditions on $\Sigma \cup \Pi$. To
this aim let us consider functions $f\in
W^{2,2}_{\mathrm{loc}}(\R^{2}\setminus (\Sigma \cup \Pi )) \cap
L^{2}$ which are continuous on $\Sigma$. For a sufficiently small
positive number $\rho$ the restriction $f\upharpoonright
_{\mathcal{C}_{\rho ,i}}$ to the circle $\mathcal{C}_{\rho
,i}\equiv \mathcal{C}_{\rho}(y_{i}):=\{q\in \R^2:|q-y^{(i)}|=\rho
\}$ is well defined. Furthermore, we will say that function $f$
belongs to $D(\dot{H}_{\alpha , \beta })$ if and only if the
following limits,
$$ \Xi_{i}(f):=-\lim _{\rho \rightarrow 0}\frac {1}{\ln \rho }f
\upharpoonright _{\mathcal{C}_{\rho ,i}}\,,\quad
\Omega_{i}(f):=\lim _{\rho \rightarrow 0}[f\upharpoonright
_{\mathcal{C}_{\rho ,i}} +\Xi_{i}(f)\ln \rho ] $$
for $i=1,\dots,n$, and
$$ \Xi_{\Sigma }(f)(x_{1}):=
\partial_{x_{2}} f(x_{1},0+)-
\partial_{x_{2}}f(x_{1},0-)\,, \quad
\Omega_{\Sigma }(f)(x_{1}):= f(x_{1},0) $$
are finite and satisfy the relations
\begin{equation}  \label{boucon}
2\pi \beta _i \Xi _{i}(f)=\Omega  _{i}(f)\,, \quad \Xi
_{\Sigma}(f)(x_{1})=-\alpha \Omega  _{\Sigma }(f)(x_{1})\,,
\end{equation}
where $\beta _{i}\in \R$. For simplicity we put $\beta \equiv
(\beta _1 ,\dots,\beta _n )$ in the following. Finally, we define
the operator $\dot H_{\alpha, \beta}:D(\dot H_{\alpha,
\beta})\rightarrow L^{2}$ acting as
$$ \dot H_{\alpha, \beta}f(x)=-\Delta f(x)\quad \mathrm{for}\quad
x\in \R^{2} \setminus (\Sigma \cup \Pi)\,. $$
The integration by parts shows that $\dot H_{\alpha, \beta}$ is
symmetric; let $H_{\alpha ,\beta }$ denote its closure. To check
that the latter is self-adjoint let us consider an auxiliary
operator $\dot{H}_{\alpha }$ defined as the Laplacian with
boundary condition (\ref{boucon}) and the additional restriction
$\Xi_{i}(f)=\Omega _{i}(f)=0$ for $f\in D(\dot{H}_{\alpha })$ and
all $i=1,...,n$. It is straightforward to see that the operator
$\dot{H}_{\alpha }$ is symmetric with deficiency indexes $(n,n)$,
and moreover, that the first equation of (\ref{boucon}) determines
$n$ symmetric linearly independent boundary conditions; thus using
the standard result \cite[Thm.~XII.30]{DS} we conclude that
$H_{\alpha ,\beta }$ is self-adjoint.

\begin{remarks}
\rm{ (a) The parameters determining the point interactions clearly
differ from the $\tilde\beta_i$ used in (\ref{forHam}), for
instance, absence of such an interaction formally means that
$\beta_i=\infty$.
\\ [.5ex] (b) With a later purpose on mind we introduce some
notations. Let $H_{\beta}:= H_{0,\beta }$ be defined as the
Laplacian with the point interactions only. Furthermore, let
$\tilde{H}_{\alpha }$ denote the Laplace operator with the point
perturbations (supported by $\Pi$) removed; this operator formally
corresponds to $H_{\alpha ,\infty}$. It is well known that both
these operators are self-adjoint -- cf.~\cite{AGHH}.}
\end{remarks}

\subsection{The resolvent}\label{resol1}

To perform spectral analysis of $H_{\alpha ,\beta }$ we will need
its resolvent. Given $z\in \rho(-\Delta)=\C\setminus [0,\infty)$
denote by $R(z)=(-\Delta -z)^{-1}$ the free resolvent, which is
well known to be an integral operator in $L^{2}$ with the kernel
\begin{equation} \label{kerneG}
G_{z}(x,x')=\frac{1}{(2\pi)^{2}}\int _{\R^{2}}\frac{\mathrm{e}^
{ip(x-x')}}{p^{2}-z}\,\mathrm{d}p=\frac{1}{2\pi}K_{0}
(\sqrt{-z}|x-x'|),
\end{equation}
where $K_{0}(\cdot)$ is the Macdonald function and the function
$z\mapsto \sqrt{z}$ has conventionally a cut at the positive
halfline. Moreover, denote by $\mathbf{R}(z)$ the unitary operator
defined as $R(z)$ but acting from $L^{2}$ to $W^{2,2}\equiv
W^{2,2}(\R^{2}).$

To construct the resolvent of $H_{\alpha, \beta}$ we will need two
auxiliary Hilbert spaces, $\mathcal{H}_0:= L^{2}(\mathbb{R})$ and
$\mathcal{H}_1:= \mathbb{C}^n$, and the corresponding trace maps
$\tau _0:W^{2,2}\to \mathcal{H}_0$ and $\tau _1 : W^{2,2}\to
\mathcal{H}_1$ which act as
 $$
 \tau _0  f:=f\!\upharpoonright_{\,\Sigma }\,, \quad \tau_1
 f:=f\!\upharpoonright_{\,\Pi}=(f\!\upharpoonright _{\,\{y^{(1)}\}},
 \dots,f\!\upharpoonright _{\,\{y^{(n)}\}})\,,
 $$
respectively; in analogy with the previous section the above
notations indicate the appropriate restrictions. By means of $\tau
_{i}$ we can define the canonical embeddings of $\mathbf{R} (z)$
to $\mathcal{H}_{i}$, i.e.
\begin{equation} \label{embedd}
\mathbf{R}_{iL}(z)=\tau _{i}R(z):L^{2}\rightarrow \mathcal{H}
_{i}\,,\quad \mathbf{R}_{Li}(z)=[\mathbf{R}_{iL}(z)]^{\ast
}:\mathcal{H} _{i}\rightarrow L^{2}\,,
\end{equation}
and
$$
\mathbf{R}_{ji}(z)=\tau _{j}\mathbf{R}_{Li}(z):\mathcal{H}%
_{i}\rightarrow \mathcal{H}_{j}\,. $$
To express the resolvent of $H_{\alpha, \beta}$ we need the
operator-valued matrix
$$
\Gamma (z)=[\Gamma
_{ij}(z)]:\mathcal{H}_{0}\oplus\mathcal{H}_{1}\rightarrow
\mathcal{H}_{0}\oplus\mathcal{H}_{1},
$$
where $\Gamma _{ij}(z):\mathcal{H}_{j}\rightarrow \mathcal{H}_{i}$
are the operators given by
 \begin{eqnarray*} \label{forg22}
 \Gamma _{ij}(z)g &:=& -\mathbf{R}_{ij}(z)g \qquad
 \mathrm{for}\quad i\neq j \quad \mathrm{and }\quad g\in
 \mathcal{H}_{j}\,, \\ \Gamma_{00}(z)f &:=& \left[\alpha^{-1}
 -\mathbf{R}_{00}(z)\right] f \qquad \mathrm{if} \quad
 f\in \mathcal{H}_0\,, \\ \Gamma _{11}(z)\varphi &:=&
 \left[ s_{\beta _{l}}(z) \delta_{kl} - G_{z}(y^{(k)},y^{(l)}) (1\!-\!
 \delta_{kl}) \right]_{k,l=1}^{n} \varphi \qquad \mathrm{for}
 \quad \varphi \in \mathcal{H}_1\,,
 \end{eqnarray*}
and $s_{\beta_{l}}(z)= \beta_{l}+s(z):=\beta
_{l}+\frac{1}{2\pi}(\ln \frac{\sqrt{z}} {2i}-\psi(1))$, where
$-\psi(1)\approx 0.577$ is the Euler number --
cf.~\cite[Sec.~I.5]{AGHH}.

We will also need the inverse of $\Gamma (z)$. To this aim let us
denote by $\mathcal{D}$ the set of $z\in \C$ such that $\Gamma
(z)$ is boundedly invertible; as we will see $\mathcal{D}$
coincides with the resolvent set of $H_{\alpha ,\beta }$. For
$z\in \mathcal{D}$ the operator $\Gamma _{00}(z)$ is invertible
and thus it makes sense to define $D(z)\equiv
D_{11}(z):\mathcal{H}_{1}\rightarrow \mathcal{H}_{1}$ by
\begin{equation} \label{Gamhat}
D(z)=\Gamma _{11}(z)-\Gamma _{10}(z)\Gamma _{00}(z)^{-1}\Gamma
_{01}(z)
\end{equation}
which is invertible for $z\in \mathcal{D}$; the above operator
will be called the \emph{reduced determinant} of $\Gamma $. By a
straightforward calculation one can check that the inverse of
$\Gamma (z)$ is given by
\begin{equation}\label{invgam}
[\Gamma(z)]^{-1}:\mathcal{H}_{0}\oplus\mathcal{H}%
_{1}\rightarrow \mathcal{H}_{0}\oplus\mathcal{H}_{1},
\end{equation}
with the "block elements" defined by
 \begin{eqnarray*}
 \left[\Gamma(z)\right]_{11}^{-1} &=& D(z)^{-1}\,, \\
 \left[\Gamma(z)\right]_{00}^{-1} &=&
 \Gamma_{10}(z)^{-1} \Gamma_{11}(z)D(z)^{-1}
 \Gamma_{10}(z)\Gamma_{00}(z)^{-1}\,, \\
 \left[\Gamma(z)\right]_{01}^{-1} &=& -\Gamma_{00}(z)^{-1}
 \Gamma _{01}(z) D(z)^{-1}\,, \\
 \left[\Gamma(z)\right]_{10}^{-1} &=& -D(z)^{-1}
 \Gamma_{10}(z)\Gamma_{00}(z)^{-1}\,;
\end{eqnarray*}
in the above formulae we use notation $\Gamma _{ij}(z)^{-1}$ for
the inverse of $\Gamma _{ij}(z)$ and $[\Gamma (z)]_{ij}^{-1}$ for
the matrix element of $[\Gamma (z)]^{-1}$.

With these preliminaries we are ready to state the sought formula
for the explicit form of the resolvent of $H_{\alpha ,\beta }$.

\begin{theorem} \label{resoth}
For any $z\in \rho (H_{\alpha ,\beta })$ with $\mathrm{Im \,}z>0$
we have
\begin{equation} \label{resolv}
R_{\alpha ,\beta }(z)\equiv (H_{\alpha ,\beta
}-z)^{-1}=R(z)+\sum_{i,j=0}^{1} \mathbf{R}
_{Li}(z)[\Gamma(z)]_{ij}^{-1}\mathbf{R}_{jL}(z)\,.
\end{equation}
\end{theorem}
\begin{proof}
We employ again the vector notation, $\Xi(f)\equiv
(\Xi_1(f),\dots,\Xi_n(f))$ and $\Omega(f)\equiv(\Omega_1(f),\dots,
\Omega_n(f))$. We have to check that $f\in D(H_{\alpha,\beta})$
holds if and only if $f=\tilde R_{\alpha,\beta}(z)g$ for some
$g\in L^2$, where $\tilde R_{\alpha,\beta}(z)$ denotes the
operator at the right-hand side of the last equation. Suppose that
$f$ is of this form. It belongs obviously to
$W_{\mathrm{loc}}^{2,2}(\mathbb{R}^{2}\setminus (\Sigma \cup \Pi))
\cap L^2$ because the same is true for all its components.
Combining the definitions of $\mathbf{R}_{ij},\:
[\Gamma(z)]_{ij}^{-1}$, and the functionals $\Xi_i$ and $\Omega_i$
introduced above with the asymptotic behaviour of Macdonald
function, specifically
\begin{equation}
\label{Macdon} K_{0}(\sqrt{-z}\rho )\rightarrow -\ln \rho -2\pi
s(z)+\mathcal{O}(\rho)\quad \mathrm{for}\quad \rho\rightarrow 0\,,
\end{equation}
we arrive at
 \begin{eqnarray*}
 2\pi \Xi (f) &\!=\!& \sum_{i=0}^1\, [\Gamma(z)]_{1i}^{-1}
 \mathbf{R}_{iL}(z)g\,, \\
 \Omega (f) &\!=\!& \mathbf{R}_{1L}(z)g -\sum_{i=0}^1\,
 \Gamma_{10}(z) [\Gamma(z)]_{0i}^{-1}\mathbf{R}_{iL}g-
 s(z)\sum_{i=0}^1\, [\Gamma(z)]_{1i}^{-1}\mathbf{R}_{iL}(z)g\,.
 \end{eqnarray*}
Let us consider separately the components of $\Xi(f),\, \Omega
(f)$ coming from the behaviour of $g$ at the points of the set
$\Pi$ and on $\Sigma $, i.e. for $i=1,2$, which means to define
the vectors $\Xi^i(f):=\frac{1}{2\pi} [\Gamma(z)]_{1i}^{-1}
\mathbf{R}_{iL}g$ and
 \begin{eqnarray*}
 \Omega^0(f) &\!:=\!& \left[- \Gamma_{10}(z)[\Gamma(z)]_{00}^{-1}
 -s(z)[\Gamma(z)]_{10}^{-1}\right] \mathbf{R}_{0L}g\,, \\
 \Omega^1(f) &\!:=\!& \left[ 1-\Gamma_{10}(z)[\Gamma(z)]_{01}^{-1}
 -s(z) [\Gamma(z)]_{11}^{-1}\right] \mathbf{R}_{1L}g\,.
 \end{eqnarray*}
Using the properties of $[\Gamma_{ij}(z)]$ and its inverse it is
straightforward to check that $\Omega^{i}_{k}(f)=2\pi \beta_{k}
\Xi^{i}_{k}(f)$ holds for $i=0,1$ and $k=1,...,n$; the symbols
$\Omega^{i}_{k}(f)$, $\Xi^{i}_{k}(f)$ mean here the $k$-th
component of $\Omega^{i}(f)$, $\Xi^{i}(f)$ respectively. Similar
calculations yield the relation $\Xi_{\Sigma }(f)= -\alpha
\Omega_{\Sigma }(f)$ which shows that $f$ belongs to
$D(H_{\alpha,\beta })$, and the converse statement, namely that
any function from $D(H_{\alpha,\beta })$ admits a representation
of the form $f=\tilde R_{\alpha,\beta}(z)g$. To conclude the
proof, observe that for a function $f\in D(H_{\alpha,\beta })$
which vanishes on $ \Sigma \cup \Pi$ we have $(-\Delta-z)f=g$.
Consequently, $\tilde R_{\alpha,\beta}(z)= R_{\alpha,\beta}(z)$ is
the resolvent of the Laplace operator in $L^2$ with the boundary
conditions (\ref{boucon}).
\end{proof}

\subsection{Another form of the resolvent }

With a later purpose on mind it is useful to look at the model in
question also from another point of view, namely as a
point-interaction perturbation of the ``line only'' Hamiltonian
$\tilde{H}_{\alpha }$. In the same way as above we can check that
the resolvent of $\tilde{H}_{\alpha }$ is the integral operator
\begin{equation}\label{resoal}
  R_{\alpha }(z)=R(z)+R_{L0}(z)\Gamma ^{-1}_{00}R_{0L}(z)
\end{equation}
for any given $z\in \rho (\tilde{H}_{\alpha })=\C\setminus
[-\frac{1}{4}\alpha ^{2},\infty )$. Define now the operators
$\mathbf{R}_{\alpha ; L1}(z):\mathcal{H}_{1}\rightarrow L^{2}$ and
$\mathbf{R}_{\alpha ;1L}(z):L^2 \rightarrow \mathcal{H}_{1}$ by
\begin{equation} \label{reRalp}
\mathbf{R}_{\alpha ; 1L}(z)f=R_{\alpha }(z)f\upharpoonright _{\Pi}
\quad \mathrm{for}\quad  f\in L^2
\end{equation}
and $\mathbf{R}_{\alpha ;L1}(z)=\mathbf{R}^{\ast }_{\alpha ;
1L}(z)$. The Hamiltonian $H_{\alpha ,\beta }$ is obtained by
adding a finite number of point perturbations to
$\tilde{H}_{\alpha }$. Consequently, the difference of the
resolvents $R_{\alpha ,\beta }$ and $R_{\alpha }$ is given by
Krein's formula
$$
  R_{\alpha ,\beta }(z)=R_{\alpha }(z)+\mathbf{R}_{\alpha ;
  L1}(z)\Gamma _{\alpha; 11}(z)^{-1}\mathbf{R}_{\alpha ;1L}(z)
$$
with
$$ \Gamma _{\alpha; 11}(z)\varphi =
 \left( s^{(\alpha)}_{\beta,k}(z) \delta_{kl} -
 G^{(\alpha)}_z(y^{(k)}, y^{(l)}) (1\!-\! \delta_{kl}) \right)
 \varphi \quad \mathrm{for} \quad \varphi \in
 \mathcal{H}_1\,, $$
where $s^{(\alpha)}_{\beta,k}(z):= \beta_k - \lim_{\eta\to 0}
\left( G^{(\alpha)}_z(y^{(k)}, y^{(k)}\!+\eta) + \frac{1}{2\pi}
\ln|\eta| \right)$ and $G^{(\alpha)}_z$ is the integral kernel of
the operator $R_\alpha(z)$. In fact, this can be simplified as
follows.

\begin{proposition} \label{resoprop}
For any $z\in \rho(H_{\alpha,\beta })$ with $\mathrm{Im\,}z>0$ we
have
 $$ R_{\alpha,\beta }(z)
 =R_{\alpha}(z) + \mathbf{R}_{\alpha;L1}(z)
 D(z)^{-1} \mathbf{R}_{\alpha;1L}(z)\,. $$
\end{proposition}
\begin{proof}
Using the asymptotic behaviour of the Macdonald function we get
 $$ s^{(\alpha)}_{\beta,k}(z) =s_{\beta_{k}}(z)
 -(\mathbf{R}_{10}(z)\Gamma_{00}(z)^{-1}
 \mathbf{R}_{01}(z))_{kk}\,.$$
This yields $\Gamma_{\alpha;11}(z) =D(z)$, and thus the claim of
the proposition.
\end{proof}


\section{Spectral analysis}
\label{Spectrum} \setcounter{equation}{0}

We begin the spectral analysis of $H_{\alpha ,\beta }$ by
localizing the essential spectrum. To this aim let us consider the
auxiliary ``line-only'' operator $ \tilde{H}_{\alpha }$ introduced
above. Separating variables and using the fact that
one-dimensional Laplace operator with a single point interaction
of coupling constant $\alpha$ has just one isolated isolated
eigenvalue equal to $-\frac{1}{4}\alpha ^{2}$ we find that $\sigma
(\tilde{H}_{\alpha })= \sigma _{\mathrm{ac}}(\tilde{H}_{\alpha
})=[-\frac{1}{4}\alpha ^{2},\infty )$. The point interactions in
$H_{\alpha,\beta}$ represent by Proposition~\ref{resoprop} a
finite-rank perturbation of the resolvent, hence the essential
spectrum is preserved by Weyl's theorem. Moreover, the explicit
expression of the resolvent makes it possible to employ
\cite[Thm.~XIII.19]{RS} to conclude that the singularly continuous
spectrum of $H_{\alpha,\beta}$ is empty, i.e. that
 \begin{equation} \label{ess-ac}
 \sigma_{\mathrm{ess}}(H_{\alpha,\beta})=
 \sigma_{\mathrm{ac}}(H_{\alpha,\beta})=
 [-\frac{1}{4}\alpha^2,\infty )\,.
 \end{equation}
To demostrate the existence of isolated points of the spectrum for
$H_{\alpha ,\beta}$ and to find the corresponding eigenvectors we
employ the following equivalences,
 \begin{eqnarray}
 z\in\sigma_{\mathrm{d}}(H_{\alpha,\beta})
 &\Leftrightarrow& 0\in\sigma_{\mathrm{d}}(\Gamma(z))\,, \;\;
 \dim\ker\Gamma(z) =\dim \ker (H_{\alpha,\beta}\!-\!z)\,,
 \phantom{AAAA} \label{eivcon} \\
 H_{\alpha,\beta}\phi_z =z\phi_z &\Leftrightarrow&
 \phi_z=\sum_{i=0}^{1}\mathbf{R}_{Li}(z) \eta_{i,z} \quad
 \mathrm{for} \; z\in\sigma_{\mathrm{disc}}(H_{\alpha,\beta})\,,
 \label{discon}
 \end{eqnarray}
where $(\eta_{0,z},\eta_{1,z})\in \ker\Gamma(z)$. They are nothing
else than a generalization of the Birman-Schwinger principle to
the situation when the interaction in the Schr\"odinger operator
in question is singular and supported by a zero-measure set; in
the present form they follow from an abstract result of
\cite[Thm.~3.4]{AP2}. Thus to investigate the discrete spectrum it
suffices to study zeros of the operator-valued function $z\mapsto
\Gamma (z)$. This will the starting point for considerations in
the rest of this section.


\subsection{Discrete spectrum for
one point interaction} \label{disspe}

We start with the simplest case when the interaction in $H_{\alpha
,\beta }$ is supported by $\Sigma $ and at a single point $y$. In
such a case, of course, we can choose $y=(0,a)$ with $a>0$ without
loss of generality. As indicated above the spectrum in
$[-\frac{1}{4}\alpha^2,\infty )$ is purely absolutely continuous;
our aim is to show that $H_{\alpha ,\beta }$ has always exactly
one isolated eigenvalue and to investigate its dependence  on the
distance $a$ between $y$ and $\Sigma $. In particular, we will
show that the eigenvalue behavior for large $a$ basically depends
on whether the number
\begin{equation} \label{xibeta}
\epsilon _{\beta }=-4\,\mathrm{e}^{2(-2\pi \beta +\psi (1))}\,,
\end{equation}
where $-\psi(1)\approx 0.577$ is the Euler number, belongs to the
absolutely continuous spectrum or not; recall that
$\epsilon_{\beta }$ is the only isolated eigenvalue of the
point-interaction Hamiltonian $H_{\beta }$ --
cf.~\cite[Sec.~I.5]{AGHH}.

Since zeros $\Gamma (z)$ determine eigenvalues of $H_{\alpha
,\beta }$, it is convenient to rewrite the operator $\Gamma (z)$
in a more explicit form. It is straightforward to see that its
part $\Gamma _{00}(z)$ acts in the momentum representation as a
simple multiplication, and therefore
$$
\Gamma _{00}(z)f(x)=\frac{1}{(2\pi )^{1/2}} \int _{\R}\left[
\frac{1}{\alpha }- \frac{i}{2(z-p^{2})^{1/2}}\right]
\tilde{f}(p)\, \mathrm{e}^{ipx}\, \mathrm{d}p\,.
$$
Moreover, using the expression for the Green function of the
one-dimensional Laplace operator,
\begin{equation} \label{1dimGf}
\frac{1}{2\pi}
\int_{\R}\frac{\mathrm{e}^{ipx}}{p^{2}-z}\,\mathrm{d}p=
\frac{i}{2\sqrt{z}}\, \mathrm{e}^{i\sqrt{z}\left| x\right|}\,,
\end{equation}
we can express the ``off-diagonal'' operator components as
\begin{equation} \label{forGij}
(\Gamma _{01}(z)\phi )(x)=\nu _{z}^{+}(x)\phi \,,\quad
\Gamma_{10}(z)f=\int_{\R}\nu ^{-} _{z}(x)f(x)\,\mathrm{d}x\,,
\end{equation}
for $\phi \in \mathcal{H}_{1}$ and $f\in \mathcal{H}_{0}$,
respectively, where
\begin{equation} \label{formuv}
\nu _{z}^{\pm}(x) := \int_{\R}v_{z}(p)\, \mathrm{e}^{\pm
ipx}\,\mathrm{d}p\,,\quad v_{z}(p):=\frac{i}{4\pi
}\frac{\mathrm{e}^{i(z-p^{2})^{1/2}a }} {(z-p^{2})^{1/2}}\,.
\end{equation}
While later we will consider analytic continuation of some of the
resolvent ``constituents'', with operators (\ref{forGij}) it is
sufficient to stay  at the first sheet of $z\mapsto
(z-p^{2})^{1/2}$, i.e. to suppose that $\mathrm{Im}\,
(z-p^{2})^{1/2}>0$. In that case the functions $\nu^{\pm}_{z}$
belong to $\mathcal{H}_{0}$, and consequently, the
``off-diagonal'' operators, $\Gamma _{ij}(z)$ with $i\neq j$, are
well defined.

To proceed further we make two observations. The first is the
equivalence
$$
0\in \sigma _{\mathrm{d}}(\Gamma (z))\Leftrightarrow 0\in \sigma
_{\mathrm{d}}(D(z))\,,
$$
where $D(z)$ is the reduced determinant of $\Gamma (z)$ given by
(\ref{Gamhat}); this means that it suffices to investigate zeros
of the map $z\mapsto D(z)$. Secondly, as we know that
$H_{\alpha,\beta}$ is self-adjoint, we can restrict ourselves to
$z=-\kappa ^2$ with $\kappa >0$. For convenience we introduce the
abreviations $\check{\Gamma} (\kappa ):=\Gamma (-\kappa ^2)$,
$\:\check{D}(\kappa )=D(-\kappa ^{2})$, and the analogous symbols
for other functions. By a straightforward computation using
formulae (\ref{formuv}), (\ref{forGij}) one can check that
$\check{D}(\kappa )$ is an operator of multiplication,
$\check{D}(\kappa )\varphi =\check{d }(\kappa )\varphi$, by the
number
$$
\check{d}(\kappa)\equiv \check{d }_{a}(\kappa):=
\check{s}_{\beta}(\kappa )-\check{\phi}_{a}(\kappa )\,,
$$
where
\begin{equation} \label{phiaka}
\check{\phi}_{a}(\kappa ):= \frac{\alpha }{4\pi }
\int_{\R}\frac{\mathrm{e}^{-2(p^{2}+\kappa ^{2})^{1/2} a }}{
(2(p^{2}+\kappa ^{2})^{1/2}-\alpha )(p^{2}+\kappa ^{2})^{1/2}}\,
\mathrm{d}p
\end{equation}
and
$$
\check{s}_{\beta }(\kappa )=s_{\beta }(-\kappa ^{2}):=\beta +
\frac{1}{2\pi }\left[ \ln \frac{\kappa }{ 2}-\psi (1)\right].
$$
Consequently, roots of the equation
\begin{equation} \label{gamche}
\check{d}_{a}(\kappa )=0 \quad \mathrm{for} \quad \kappa \in
(\alpha /2 ,\infty)
\end{equation}
determine through $z=-\kappa ^2$ the discrete spectrum of
$H_{\alpha ,\beta }$.

Now we are ready to state a claim which characterizes the discrete
spectrum of $H_{\alpha ,\beta }$ in case of a single point
perturbation.

\begin{theorem} \label{isospe}
For given $\alpha >0$ and $\beta \in \R$ the operator $H_{\alpha
,\beta }$ has exactly one isolated spectrum $-\kappa_{a}^2$ with
the eigenvector which can be represented by
\begin{equation} \label{eigfu1}
\mathrm{const }\int_{\R^{2}}\left(
\frac{\mathrm{e}^{-ip_{2}a}}{2\pi}+\frac{\alpha\,
\mathrm{e}^{-(p_{1}^{2}+ \kappa _{a}^{2})^{1/2} a
}}{2(p_{1}^{2}+\kappa _{a}^{2})^{1/2}-\alpha }\right)
\frac{e^{ipx}}{p^{2}+\kappa _{a}^{2}}\, \mathrm{d}p\,,
\label{eigvec}
\end{equation}
where we integrate with respect to $p=(p_{1},p_{2})$.
\end{theorem}

\noindent \begin{proof} To check that there is a $\kappa _{a}$
satisfying (\ref{gamche}) it suffices to investigate the behavior
of $\check{d }_{a}$ at infinity and near the number
$\frac{1}{2}\alpha$. Using the above definitions of $\check
{s}_{\beta }$ and $\check{\phi}_{a}$ it is easy to see that the
function $\kappa \mapsto \check{d} _{a}(\kappa )$ is strictly
increasing with the limits $\check{d}_{a}(\kappa )\to \pm\infty$
as $\kappa \to\infty$ and $\kappa \rightarrow \frac{1}{2}\alpha+$,
respectively. Thus there is exactly one $\kappa _{a}\in
(\frac{1}{2}\alpha,\infty )$ such that $\check{d}_{a}(\kappa
_{a})=0$. Formula (\ref{eigfu1}) can be obtained directly from
(\ref{discon}).
\end{proof}
\vspace{1em}

Next we want to look at the asymptotic behavior of the eigenvalue
the existence of which we have just established for large as well
as small distance $a$; in this respect it is convenient to use the
notation $H_{\alpha ,\beta ,a}$ for the operator in question. The
answer is again contained in the behavior of the functions
$\check{s}_{\beta }(\cdot ),$ and $\check{\phi}_{a}(\cdot )$.
Given $\kappa \in (\frac{1}{2}\alpha ,\infty )$ we define the
function $a\mapsto \tilde{\phi}_{\kappa }(a)=\check{\phi}_{a}
(\kappa )$; using (\ref{phiaka}) it is easy to see that it is
decreasing on the indicated interval. Combining this with the fact
that $\check{s}_{\beta }(\cdot )$ is increasing we come to the
conclusion that the function $a\mapsto\kappa _{a}$ is decreasing
on $(0,\infty)$. To determine its behavior at the endpoints if the
interval let us notice that
$$
\lim_{a\rightarrow \infty }\tilde{\phi}_{\kappa }(a)=0\,.
$$
This limit in combination with the relation $\check{s}_{\beta
}(\sqrt{-\epsilon _{\beta }})=0$, where $\epsilon _{\beta }$ is
the point-interaction eigenvalue given by (\ref{xibeta}), yields
$$
\lim_{a\rightarrow \infty }\kappa_a=\sqrt{-\epsilon _{\beta
}}\quad \mathrm{if} \quad \sqrt{-\epsilon _{\beta }}\in (\alpha/2
,\infty )
$$
and
$$
\lim_{a\rightarrow \infty }\kappa_a=\frac{\alpha }{2} \quad
\mathrm{if} \quad \sqrt{-\epsilon _{\beta }}\in (-\infty
,\alpha/2]\,.
$$
Let us turn next to the behavior $a\mapsto \kappa_a$ for small
$a$. To this aim we note that for a fixed $\kappa $ the expression
$$
\check{\phi}_{0}(\kappa ):=\frac{\alpha }{4\pi }\int_{\R}\frac{1}{
(2(p^{2}+\kappa ^{2})^{1/2}-\alpha )(p^{2}+\kappa ^{2})^{1/2}}\,
\mathrm{d}p
$$
provides an upper bound for $\check{\phi}_{a}(\kappa )$. It
straightforward to check that $\check{\phi}_{0}(\kappa
)\rightarrow 0$ as $\kappa \rightarrow \infty$ and
$\check{\phi}_{0}(\kappa )\rightarrow \infty$ as $\kappa
\rightarrow \frac{1}{2}\alpha+$. It follows that there is a number
$\kappa_{0}\in (\frac{1}{2}\alpha ,\infty )$ which is a solution
of $\check{s}_{\beta }(\kappa )- \check{\phi}_{0}(\kappa )=0$ and
provides an upper bound to the function $ a\mapsto\kappa_a$. These
considerations can be summarized as follows:
\begin{theorem} \label{asymev}
The eigenvalue $-\kappa_{a}^{2}$ of $H_{\alpha ,\beta ,a}$ is
increasing as a function of the distance $a$. Moreover, we have
$$-\lim_{a\rightarrow \infty
}\kappa _{a}^{2}=\epsilon _{\beta }\quad \mathrm{if} \quad
\epsilon _{\beta }\in (-\infty ,-\frac{1}{4}\alpha ^{2}] $$
and
$$-\lim_{a\rightarrow \infty }
\kappa _{a}^{2}=-\frac{1}{4}\alpha ^{2} \quad \mathrm{if }\quad
\epsilon _{\beta } \in (-\frac{1}{4}\alpha ^{2},\infty )\,.$$
On the other hand, $-\kappa _{0}^{2}$ is the best lower bound for
$-\kappa _{a}^{2}$, i.e. we have
$$-\lim_{a\rightarrow 0 } \kappa _{a}^{2}=-\kappa _{0}^{2}\,.$$
\end{theorem}


\subsection{A mirror-symmetric pair of point interactions} \label{dispsy}

Generally speaking the case of $n=2$ can be treated within the
discussion of the discrete spectrum of $H_{\alpha,\beta}$ with
$n>1$ presented in the next subsection. Here we single out the
situation where the system has a mirror symmetry to illustrate
that it can give rise to eigenvalues embedded in the continuous
spectrum. To be specific, we assume that the interaction sites are
located symmetrically with respect to the line $\Sigma$, i.e.
$x_{1}=(0,a)$, $x_{2}=(0,-a)$ with some $a>0$, and moreover, the
coupling strengths are the same, $\beta_1=\beta_2=\beta$.

As in the case $n=1$ the relation between the number
$-\frac{1}{4}\alpha^{2}$ and the point-interaction eigenvalues
will be important for spectral properties. Consider the system
with the line component of the interaction removed which is
described by the operator $H_{\beta}$. It has
$\sigma_{\mathrm{ac}}(H_{\beta}) =[0,\infty)$ and at least one and
at most two eigenvalues. Let us denote them $\mu _1$, $\mu _2$ and
assume that $\mu _1<\mu _{2}$; if there exists only one eigenvalue
we put $\mu _{2}:=0$. From the explicit resolvent formula
\cite[Sec.~II.4]{AGHH} it follows that $\mu_{i}=-\kappa _{i}^{2}
$, where $\kappa _{i}$ are solutions of the equation
$$ \check{s}_{\beta }(\kappa )^2-K_{0}(2\kappa a)^2=0\,, \quad
\kappa>0\,, $$
which implies the inequalities
\begin{equation}\label{ineqmu}
\mu_1<\epsilon_{\beta } <\mu _{2}\,;
\end{equation}
they follow also from Dirichlet-Neumann bracketing
\cite[Sec.~XIII.15]{RS} and it is useful to note that the number
$\mu_{1}\,,\mu_{2}$ is the eigenvalue corresponding to the
symmetric and antisymmetric eigenfunction of $H_{\beta }$,
respectively.

To find the isolated eigenvalue of $H_{\alpha , \beta}$ we will
employ the BS-principle expressed by (\ref{eivcon}). Proceeding
similarly as in the previous section we show that the number
$-\tilde {\kappa }^2$ is an eigenvalue of $H_{\alpha ,\beta }$
\emph{iff} $\tilde{\kappa }$ is a solution of
\begin{equation}\label{gamma2}
  \check{d }(\kappa )=0\quad \mathrm{for}\quad \kappa \in
  (\alpha/2,\infty)\,,
\end{equation}
where the function $\check{d}(\cdot)$ means the determinant of
$\check{D}(\cdot)$ being thus given by
$$
\check{d}(\kappa )=(\check{s}_{\beta }(\kappa )+K_{0}(2\kappa
a))(\check{s}_{\beta }(\kappa )-K_{0}(2\kappa
a)-2\check{\phi}_{a}(\kappa ))
$$
and $\check{d}(\kappa )$ is again given by (\ref{phiaka}), i.e.
\begin{equation} \label{phia2}
\check{\phi}_{a}(\kappa )=\frac{\alpha }{4\pi }
\int_{\R}\frac{\mathrm{e}^{-2(p^{2}+\kappa ^{2})^{1/2}a }}{
(2(p^{2}+\kappa ^{2})^{1/2}-\alpha )(p^{2}+\kappa ^{2})^{1/2}}\,
\mathrm{d}p\,.
\end{equation}
\smallskip

Now we can describe the point spectrum of $H_{\alpha ,\beta}$ in
the given situation.
\begin{theorem} \label{theign}
$H_{\alpha,\beta}$ has always at least one isolated eigenvalue.
Moreover, \\ [.5ex] (i) if $-\frac{1}{4}\alpha^2 <\mu_2 <0$, then
$H_{\alpha ,\beta }$ has one isolated eigenvalue and one embedded
eigenvalue which is equal to $\mu_2$, \\ [.5ex] (ii) on the other
hand, if $\mu_2 <-\frac{1}{4}\alpha^2$, then $H_{\alpha ,\beta }$
has two isolated eigenvalues the larger of which is given by
$\mu_2$.
\end{theorem}
\begin{proof}
Using the behavior of functions $\check{s }_{\beta },\, K_{0}$,
and $\check{\phi}_{a}$ at infinity and near the number
$\frac{1}{2}\alpha$ we can conclude that the equation
$$
 \check{s}_{\beta }(\kappa)-K_{0}(2\kappa a)
 -2\check{\phi}_{a}(\kappa )=0
$$
coming from the second factor in the spectral condition has for
any parameter values exactly one solution in $(\frac{1}{2}\alpha,
\infty)$ which naturally solves also (\ref{gamma2}); this means
that the operator $H_{\alpha ,\beta }$ has always at least one
isolated eigenvalue. Moreover, if $\mu_2 <-\frac{1}{4}\alpha^2$
the equation (\ref{gamma2}) has one more solution given by the
number $\kappa _{2}$; this completes the proof of (ii). Assume
next $-\frac{1}{4}\alpha^2 <\mu_2 <0$. As we have already
mentioned the number $\mu_{2}$ is the eigenvalue of $H_{\beta }$
corresponding to eigenfunction $\psi _{\mu_{2}}$ antisymmetric
w.r.t. $\Sigma$. It is easy to see that $\psi _{\mu_{2}}\in
D(H_{\alpha ,\beta })$ and both the boundary functions $\Xi
_{\Sigma }(\psi _{\mu _{2}})\,,\Omega _{\Sigma }(\psi _{\mu
_{2}})$ vanish. This implies $H_{\alpha ,\beta }\psi _{\mu
_{2}}=H_{\beta }\psi _{\mu _{2}}$, in other words that $\psi _{\mu
_{2}}$ is at the same time an eigenvector of $H_{\alpha ,\beta }$
corresponding to $\mu _{2}$.
\end{proof}
\begin{remark}
{\rm Let us note here that the condition
\begin{equation}\label{xi>thr}
\epsilon_{\beta }>-\frac{1}{4}\alpha^2
\end{equation}
is sufficient for $\mu_{2}>-\frac{1}{4}\alpha^2$ in view of
(\ref{ineqmu}), while the converse statement is not true in
general. It may happen when the distance $a$ is sufficiently small
that even if $\epsilon_{\beta }$ is below the threshold of the
essential spectrum, the number $\mu_2$ would satisfy $\mu_{2}>
-\frac{1}{4}\alpha^2$ so according to Theorem~\ref{theign} it will
appear in spectrum of $H_{\alpha ,\beta  }$ as an embedded
eigenvalue. }
\end{remark}


\subsection{Finitely many point interactions}

Let us finally turn to analysis of the discrete spectrum in the
general case with finitely many points of interaction and coupling
constants determined by components of the vector $\beta =(\beta_1,
\dots,\beta_n)$. We assume that the perturbations are located at
$y^{(i)}=(l_{i},a_{i})$, where $l_{i}\in \R$, $a_{i}\in \R
\setminus \{0\}$, and denote by
$$ d_{ij}:= |y^{(i)}-y^{(j)}| $$
the distances between them. Our strategy will be similar as
before, namely, to recover the discrete spectrum of $H_{\alpha
,\beta }$ we will employ the equivalence (\ref{eivcon}) which
allows us to describe eigenvalues of $H_{\alpha ,\beta }$ in the
terms of zeros of $z \mapsto \Gamma (z)$. This in turn can be
reduced to the problem of finding zeros of the $n\times n$ matrix
$D(z):=\Gamma _{11}(z)-\Gamma _{10}(z)\Gamma _{00}(z)^{-1}\Gamma
_{01}(z):\mathcal{H}_{1}\to \mathcal{H}_{1}$ for $z$ negative. To
proceed further we introduce the notation $\Gamma _{j;0}$ for the
$j$-th component of $\Gamma _{10}$ and $\Gamma _{i;j}$ for the
corresponding matrix element of $\Gamma _{11}$. We also introduce
the following auxiliary functions of $z\in \C \setminus
[-\frac{1}{4}\alpha^2,\infty)$,
 $$ \Theta ^j_{i_{1}}:=\Gamma
_{j;0}\Gamma ^{-1}_{00}\Gamma _{0;i_1}\,, $$
 $$ A^j_{i_2,...,i_k}:=
 \left\{
 \begin{array}{cc}
 \Gamma _{1;i_2} \dots\Gamma _{j-1;i_{j}}\,\Gamma
 _{j+1;i_{j+1}}\dots\Gamma
 _{k;i_{k}} &\quad \mathrm{if}\: j>1\,, \\
 \Gamma _{2;i_{2}}\dots\Gamma _{k;i_k} & \quad
 \mathrm{if}\: j=1\,.
 \end{array} \right.
$$
A straightforward computation shows that the determinant of
$D(\cdot)$ is given by the function $d(\cdot )$ with the values
\begin{equation}\label{determ}
d(z)=\sum _{\pi \in \mathcal{P}_{n}}\mathrm{sgn}\,\pi \left ( \sum
_{j=1}^{n}(-1)^{j} S^{j}_{p_1,...,p_n}+\Gamma _{1;p_1}\dots \Gamma
_{n;p_n}\right)(z)\,,
\end{equation}
where $ S^{j}_{p_1,...,p_n}:=\Theta ^j_{p_{1}}A^j_{p_2,...,p_n}$,
$\:\mathcal{P}_{n}$ is the permutation group of $(1,...,n)$, and
$\pi= (p_1,...,p_n)$ is an element of $\mathcal{P}_{n}$. Since we
are interested in the negative part of spectrum we put
$\check{d}(\kappa )=d(-\kappa ^2)$ and same convention will be
kept for the other expressions. According to the above general
discussion the eigenvalues of $H_{\alpha ,\beta }$ are determined
by solution of the equation
\begin{equation}\label{eigenn}
  \check{d}(\kappa )=0 \quad \mathrm{for }\quad \kappa \in
  (\alpha/2,\infty )\,.
\end{equation}
To concretize the function $\check{d}(\cdot )$ we need more
information about the functions involved in the definition of
$D(\cdot)$. We have
\begin{equation} \label{Theta}
\check{\Theta }_{k}^{j}(\kappa ) =\frac{\alpha }{4\pi
}\int_{\R}\frac{ \mathrm{e}^{-(p^{2}+\kappa ^2 )^{1/2}(|a_i |+
|a_j|)}}{(2(p^{2}+\kappa ^2 )^{1/2}-\alpha )(p^{2}+\kappa ^2
)^{1/2}}\, \e^{ip(l_{j}-l_{k})}\, \mathrm{d}p
\end{equation}
and
\begin{equation} \label{Gammjk}
\check{\Gamma }_{j;k}(\kappa )=-\frac{1}{2\pi }K_{0}(d_{jk}\kappa)
\quad \mathrm{for}\: j\neq k\,;
\end{equation}
recall that the diagonal elements for $j\geq 1$ are given by the
numbers $ \check{\Gamma }_{j;j}(\kappa)
=\check{s}_{\beta_{j}}(\kappa)$. After these preliminaries we
ready to prove the following theorem.
\begin{theorem} \label{evn-ca}
Let $\beta=(\beta  _{1},...,\beta  _{n})$, where $\beta _{i}\in
\R$ and $\alpha >0$. The operator $H_{\alpha ,\beta }$ has at
least one isolated eigenvalue and at most $n$ of them; they are
determined by solutions of the equation (\ref{eigenn}). In
particular, if all the numbers $-\beta _{i}$ are sufficiently
large then $H_{\alpha ,\beta }$ has exactly $n$ eigenvalues.
\end{theorem}
\begin{proof}
Let us consider again the operator $\dot{H}_{\alpha }$ defined in
sec.~\ref{defHam}. Since it is symmetric with deficiency indices
$(n,n)$ and $\dot{H}_{\alpha }\geq -\frac{1}{4}\alpha^2$, there
are at most $n$ eigenvalues of $H_{\alpha ,\beta }$ --
cf.~\cite[Sec.~8.4]{We}. The remaining part of the proof will be
divided into four steps.

\emph{1st step:} We will show that if all the numbers $\beta _{i}$
are sufficiently large then the equation (\ref{eigenn}) has at
least one solution. To this aim we shall investigate the behaviour
of $\check{d}(\cdot )$ at infinity and near the number
$\frac{1}{2}\alpha$. It is easy to see that for large values of
the argument $\kappa $ the behaviour of the function $\check{d}$
is determined by the term $\prod _{i=1}^{n}\check{\Gamma
}_{i;i}=\prod _{i=1}^{n}\check{s}_{\beta _{i}}$; this implies
\begin{equation}\label{auxcon}
  \check{d}(\kappa )\to \infty \quad \mathrm{as}\quad
  \kappa \to \infty \,.
\end{equation}
On the other hand, the function $\check{d}$ has a singularity at
$\frac{1}{2}\alpha$ induced by $\check {\Theta}_{j}^{i}$. This
fact allows us to conclude that if all the numbers $\beta _{i}$
are sufficiently large then the behaviour of $\check{d}(\kappa)$
near $\frac{1}{2}\alpha$ is dominated by the components of  $
-S^{j}_{j,1,2,...,j-1,j+1,...,n}$ which look like $-\check{\Theta
} _{j}^{j}\beta_{1}\dots\beta _{j-1}\,\beta _{j+1}\dots \beta
_{n}$. Since they are all negative under our assumption, we arrive
at
$$
  \check{d}(\kappa )\rightarrow -\infty \quad \mathrm{as}\quad
  \kappa \rightarrow \frac{1}{2}\alpha\,.
$$
Combining this with (\ref{auxcon}) we demonstrate the existence of
at least one solution of (\ref{eigenn}) if the coupling constants
$\beta_{i}$ are sufficiently large.

\emph{2 step:} Notice further that the functions $\check {\Gamma }
_{i;i}=\check {s}_{\beta _{i}}$ are increasing with respect to
each parameter $\beta _{i}$ while the other matrix elements of
$\check {\Gamma }$ are independent of all the $\beta _{i}$.
Combining this with the minimax principle and the results obtained
in the previous step we find that for all $\beta _{1},...,\beta
_{n}$ there exists at least one solution of (\ref{eigenn}), and
consequently, an eigenvalue of $H_{\alpha ,\beta }$.

\emph{3 step:} Let $\tilde {\kappa } $ be a solution to
(\ref{eigenn}). From (\ref{Theta}), (\ref{Gammjk}) in combination
with the explicit expression for $\check{s}_{\beta _{i}}$ one
finds that if all the coupling constants $\beta _{i}\to -\infty $
then $\tilde {\kappa }$ tends to infinity or to the number
$\frac{1}{2}\alpha$. However, the latter is excluded by the
monotonicity proved in the previous step. Thus we obtain
$$ \tilde{\kappa }\rightarrow \infty \quad \mathrm{as }\quad \beta
_{i}\rightarrow -\infty \quad \mathrm{for }\quad i=1,\dots,n\,. $$

\emph{4 step:} Using the explicit formulae for operators
$\check{\Gamma }_{i,j}$ one check that operator $\check{\Gamma
}(\kappa )$ approaches $\check{S}(\kappa )$ in the norm operator
sense as $\kappa \to \infty$, where $\check{S}(\kappa )$ is the
operator-valued diagonal matrix given by
\begin{eqnarray*}
\check{S}_{00}(\kappa )&=& 0\,,\\ \check{S}_{11}(\kappa )&=&
[\check{s}_{\beta _{k}}(\kappa )\delta _{kl }]_{k,l=1}^{n}\,,\\
\check{S}_{ij}(\kappa )&=& 0 \quad \mathrm{for} \quad i,j =0,1
\quad \mathrm{and} \quad i\neq j\,.
\end{eqnarray*}
Since there exist $n$ solutions of the operator equation
$\check{S}(\kappa )=0$ we arrive at the final conclusion that for
$-\beta _{i}$ sufficiently large the operator $H_{\alpha ,\beta }$
has the ``full number'' $n$ of isolated eigenvalues.
\end{proof}


\section{Resonances}
\label{resona} \setcounter{equation}{0}

Determining the spectrum as a set does not exhaust interesting
properties of the present model; now we turn to features
``hidden'' in the continuous component (\ref{ess-ac}). We will
concentrate at the negative part of this interval, where in the
absence of point perturbations we have a simple one-dimensional
transport: the wavefunctions factorize into the transverse factor
which is the eigenfuction of the one-dimensional point
interaction, and the longitudinal one which a wave packet which
moves and spreads in the usual way. If we add now point
perturbation(s) the transport may be affected by tunneling between
the line and these singular ``potential wells'', at least if such
a process is energetically allowed; our goal stated in the
introduction is to show existence of ``tunneling'' resonances and
to find their properties. For the sake of simplicity we shall
consider mostly (with the exception of Sec.~\ref{broken} below)
the case when the Hamiltonian $H_{\alpha,\beta}$ has a single
point perturbation.

Following the standard ideology, to find resonances we have to
construct the analytical continuation of $z\mapsto R_{\alpha
,\beta }(z)$ to the second sheet across the cut corresponding to
the continuous spectrum and to find poles of this continuation.
Our main insight is that the constituents of the operator at the
right-hand side of (\ref{resolv}) can be separately continued
analytically, and moreover as we remarked above, for the factors
(\ref{forGij}) in fact no continuation is needed, i.e. we may
suppose that $\mathrm{Im}\, (z-p^{2})^{1/2}>0$. Thus we have to
deal only with the middle factor in the interaction term of
(\ref{resolv}), in other words, we can extend the Birman-Schwinger
principle to the complex region and to look for zeros in the
analytic continuation of $D(\cdot)$. Taking into account the
structure of the auxiliary space $\mathcal{H}_0\oplus
\mathcal{H}_1$ we get in this way a problem reminiscent of the
Friedrichs model -- cf.~\cite{Fr}, or \cite[Sec.~3.2]{Ex1} for a
review.


\subsection{Resonance for $H_{\alpha,\beta}$ with a single point
interaction}\label{resec1}

The Friedrichs model analogy suggests to treat our problem
perturbatively assuming that in the ``decoupled'' case which
corresponds here to the limit $a\to \infty$ we have the point
interaction eigenvalue $\epsilon_{\beta}$ embedded in the
continuous spectrum. Following the above sketched program we
notice first that by formulae (\ref{1dimGf}), (\ref{forGij}), and
(\ref{formuv}) the operator-valued function $z\mapsto D(z)$, $z\in
\C\setminus [-\frac{1}{4}\alpha,\infty )$ is now one-dimensional,
i.e. a multiplication by the function
\begin{equation}\label{resdet}
d_{a}(z)=s_{\beta}(z)-\phi_{a}(z)\,,\quad \mathrm{where}\quad
\phi_{a}(z):=\int_{0}^{\infty }\frac{\mu (z,t)}{t-z-
\frac{1}{4}\alpha^2}\,\mathrm{d}t\,,
\end{equation}
and
$$ \mu (z,t):=\frac{i\alpha }{2^5\pi }\, \frac{(\alpha
-2i(z-t)^{1/2})\, \mathrm{e}^{2i(z-t)^{1/2}
a}}{t^{1/2}(z-t)^{1/2}}\,.$$
Since the numbers $0$ and $-\frac{1}{4}\alpha^2$ are branching
points of the function $d_{a}$ we will construct its continuation
across the interval $(-\frac{1}{4}\alpha^2,0)$ to a subset
$\Omega_-$ of the lower halfplane. Let us first consider the
second component $\phi_{a}$. To find its analytical continuation
to the second sheet for $\lambda \in (-\frac{1}{4}\alpha ^2,0) $
we define
$$\mu^{0}(\lambda,t):= \lim_{\varepsilon\to 0+}
\mu(\lambda\!+\!i\varepsilon,t) \quad \mathrm{and }\quad
I(\lambda):=\mathcal{P}\int_0^{\infty}
 \frac{\mu^{0}(\lambda,t)}{t-\lambda
 -\frac{1}{4}\alpha^2}\, \mathrm{d}t$$
with the integral understood in the principal-value sense. We also
denote
 $$ g_{\alpha ,a}(z):= \frac{i\alpha}{8}\,
 \frac{\mathrm{e}^{-\alpha a}}
 {(z +\frac{1}{4}\alpha^2)^{1/2}} \quad \mathrm{for}
 \;\; z\in\Omega_- \cup (-\frac{1}{4}\alpha ^2,0)\,; $$
then we are ready to formulate a lemma describing the analytic
continuation of $\phi_{a}$; we postpone its proof to the appendix.
\begin{lemma} \label{anacon}
The function $z\mapsto \phi _{a}(z)$ defined in (\ref{resdet}) can
be continued analytically across $(-\frac{1}{4}\alpha^2,0)$ to a
region $\Omega_-$ of the second sheet as follows,
 \begin{eqnarray*}
 \phi_a^{0}(\lambda) =I(\lambda) +g_{\alpha,a}(\lambda )
 \quad &\mathrm{for}\;& \lambda \in
 (-\frac{1}{4}\alpha^2,0)\,, \\
 \phi_a^{-}(z)= -\int_0^{\infty } \frac{\mu(z,t)}
 {t-z-\frac{1}{4}\alpha^2}\, \mathrm{d}t -2g_{\alpha,a}(z)
 \quad &\mathrm{for}\;& z\in\Omega_-\,,\: \mathrm{Im\,}z<0\,.
 \end{eqnarray*}
\end{lemma}
\bigskip

\noindent Notice that apart of fixing a part of its boundary, we
have imposed no restrictions on the shape of $\Omega_-$. The lemma
allows us to construct the sought analytic continuation of
$d_a(\cdot)$ across the indicated segment of the real axis because
the other component has no cut there. It is given by the function
$\eta_a: M\mapsto\mathbb{C}$, where $M:=\{z:\mathrm{Im\,}z>0\}\cup
(-\frac{1}{4}\alpha^2,0) \cup\Omega_-$, acting as
 $$ \eta_a(z)=s_\beta(z)-\phi_a^{l(z)}(z)\,,$$
where $l(z)=\pm$ if $\pm\mathrm{Im\,} z>0$ and $l(z)=0$ if
$z\in(-\frac{1}{4}\alpha^2,0)$, respectively; we also put
$\phi^{+}_{a}\equiv \phi _{a}$. The problem at hand is now to show
that $\eta_a(\cdot)$ has a second-sheet zero, i.e. $\eta_a(z)=0$
for some $z\in \Omega_{-}$. To proceed further it is convenient to
put $\varsigma_{\beta} :=\sqrt{-\epsilon_{\beta}}$, and since we
are interested here primarily in large distances $a$, to make the
following reparametrization,
 $$ b:=\mathrm{e}^{- a \varsigma _{\beta }}\quad \mathrm{and}\quad
 \tilde\eta(b,z):=\eta_a(z):\:[0,\infty) \times M\mapsto
 \mathbb{C}\,;
 $$
we look then for zeros of the function $\tilde\eta$ for small
values of $b$. With this notation we have
\begin{equation}\label{reparb}
\mu^{0}(\lambda,t) =\frac{\alpha }{2^5\pi }\frac{(\alpha
+2(t-\lambda )^{1/2})\, b^{2(t-\lambda )^{1/2}/\varsigma_{\beta}
}}{t^{1/2}(t-\lambda )^{1/2}}\,, \quad g_{\alpha ,a(b)}(\lambda)=
\frac{i\alpha}{8}\, \frac{b^{\alpha /\varsigma _{\beta }}}
 {(\lambda +\frac{1}{4}\alpha^2)^{1/2}}\,,
\end{equation}
for $\lambda \in (-\frac{1}{4}\alpha ^2,0)$, where $a(b):=
-\frac{1}{\varsigma_\beta}\ln b$, and similarly for the other
constituents of $\tilde\eta$. This yields our main result in this
section.
\begin{theorem} \label{resonth}
Assume that $\epsilon_{\beta }> -\frac{1}{4}\alpha^2$. For any $b$
small enough the function $\tilde\eta(\cdot,\cdot)$ has a zero at
a point $z(b)\in \Omega_-$ with the real and imaginary part,
$z(b)=\mu(b)+ i\nu(b),\; \nu(b)<0,$ which in the limit $b\to 0$,
i.e. $a\to\infty$, behave in the following way,
 \begin{equation}\label{assyRI}
 \mu(b)=\epsilon_{\beta}+\mathcal{O}(b)\,, \quad \nu(b)
 =\mathcal{O}(b)\,.
 \end{equation}
 \end{theorem}
\begin{proof} By assumption we have $\varsigma _{\beta }\in
(0,{1\over 2}\alpha)$. Using formulae (\ref{reparb}) together with
the similar expressions of $\mu (z,t)$ and $g_{\alpha ,a}(z)$ in
terms of $b$ one can check that for a fixed $b\in [0,\infty )$ the
function $\tilde\eta (b,\cdot )$ is analytic in $M$, while with
respect to both variables $\tilde\eta $ is just of the $C^1$ class
in a neighbourhood of the point $(0,\epsilon_{\beta })$. Moreover,
it is easy to see that for $\lambda $ close to $\epsilon_{\beta }$
the function $\phi^{0}_{a(b)}(\cdot)$ can majorized by the
expression $C\,b^{M}$, where $C\,,M$ are constants and $M>1$. This
implies $\tilde\eta(0,\epsilon_{\beta})=0$ and
$\partial_z\tilde\eta(0,\epsilon_{\beta}) \neq 0$. Thus by the
implicit function theorem there exists a neighbourhood $U_{0}$ of
zero and a unique function $z(b):\:U_{0}\mapsto \mathbb{C}$ such
that $\tilde\eta(b,z(b))=0$ holds for all $b\in U_{0}$. Since
$H_{\alpha,\beta}$ is self-adjoint, $\nu(b)$ cannot be positive,
while $z(b)\in (-{1\over 4}\alpha ^2,0)$ for $b\neq 0$ can be
excluded by inspecting the explicit form of $\tilde\eta $.
Finally, by the smoothness properties of $\tilde\eta$ both the
real and imaginary part of $z(b)$ are of the $C^1$ class which
yields the behaviour (\ref{assyRI}).
\end{proof}

\medskip

\begin{remark} \rm{The above theorem confirms what one expects
about the behaviour of the pole using the heuristic idea about
tunneling between the point and the line, namely that the
resonance width $\Gamma (b)=2\nu (b)$ is exponentially small for
$a$ large. It also natural to ask how the resonance pole behaves
for a general $a$, in particular, whether it may disappear for
$a\to 0$. Using the explicit formulae of Lemma~\ref{anacon} one
can check the following convergence,
$$|\phi^{-}_{a}(z)|\to 0 \quad \mathrm{as} \quad \mathrm{Im}\,z \to
-\infty\,, $$
uniformly with respect to $a$. On the other hand, it is easy to
see that
$$ |s_{\beta }(z)|\to \infty \quad  \mathrm{as} \quad \mathrm{Im}\,z
\to -\infty \,. $$
This means that the imaginary part of $z(a)$ which represents the
solution of $s_{\beta }(z)-\phi^{-}_{a}(z)=0$ is a function
uniformly bounded with respect to $a$; thus the resonance pole
survives the limit $a\to 0$.}
\end{remark}


\subsection{Resonances induced by broken symmetry} \label{broken}

If there is more than one point interaction our model may exhibit
another sort of resonances coming from broken symmetry. We
restrict ourselves to the simplest case $n=2$. As have seen in
Sec.~\ref{dispsy} the system with two points interactions placed
symmetrically with respect to the line $\Sigma $ and with equal
coupling constants $\beta _{1},\beta _{2}$ may have an embedded
eigenvalue for appropriate parameter values. If we break the
symmetry the corresponding resolvent pole will leave the
continuous spectrum and shift to the second sheet of the
analytically continued resolvent giving rise to a resonance. Of
course, there are various ways how the mirror symmetry can be
broken.


\subsubsection{Symmetry broken by a coupling constant}

Suppose first that the geometrical symmetry remains preserved,
i.e. the point interactions are located at $x_1=(0,a),\,
x_2=(0,-a)$ with $a>0$. The symmetry breaking will be due to
unequal coupling parameters: assume that the latter are $\beta
\equiv \beta _1$ and $\beta _2 =\beta +q$, where $q \in
\R\setminus \{0\}$. To get a nontrivial result, similarly as in
Sec.~\ref{dispsy} we suppose that $0>\mu_2>-\frac{1}{4}\alpha^2$.

To find the pole position we proceed as in Sec.~\ref{dispsy}; we
write the corresponding $2\times 2$ reduced determinant, construct
its analytical continuation and look for its zeros at the second
sheet. This leads to the following equation,
 \begin{equation} \label{symres}
 \eta _{q}(z)=0\,,
 \end{equation}
where
\begin{eqnarray*}
 \eta _{q}(z) &\!:=\!&  s_{\beta }(z)(s_{\beta}(z)+q)
 -K_{0}(2a\sqrt{-z})^2 \\ && -(2s_{\beta }(z) +q)\phi
^{l(z)}_{a}(z)- 2K_{0}(2a\sqrt{-z})\phi ^{l(z)}_{a}(z)
\end{eqnarray*}
and $\phi ^{l(z)}_{a}(\cdot)$ has been defined in
Lemma~\ref{anacon}. Our aim is to show that the function
$\tilde{\eta }(q,z):\R \setminus \{0\} \times M \to \C$ defined by
$ \tilde{\eta }(q,z)= \eta _{q}(z)$ has a zero in the lower
halfplane; the set $M$ is determined here as before, namely
$M=\{z: \mathrm{Im}\,z>0\}\cup \Omega _{-}$. Moreover, we put
$$\tilde{g}(\lambda):= -ig_{\alpha ,a }(\lambda )=
\frac{\alpha }{8}\, \frac{\mathrm{e}^{-\alpha a }}{ (\lambda
+\frac{1}{4}\alpha^{2})^{1/2}}$$
and use again $\kappa _{2}:=\sqrt{- \mu _{2}}$. It is also
convenient to denote
$$ \vartheta \equiv \vartheta (\kappa _{2}):=\frac{\kappa
_{2}}{\check{s} '_{\beta }(\kappa _{2})+2aK_{0} '(2a\kappa _{2})},
$$
where the primes stand for the derivatives of the corresponding
functions; with this notations we can make the following claim.
\begin{theorem}
Suppose that $\mu_2\in(-\frac{1}{4}\alpha^2,0)$, then for all
nonzero $q$ small enough the equation (\ref{symres}) has a
solution $z(q)\in \Omega_-$ with the real and imaginary part,
$z(q)=\hat{\mu }(q)+ i\hat{\nu }(q)$, which are real-analytic
functions of $q$ having the following expansions,
 \begin{eqnarray*}
 \hat{\mu}(q) &=& \mu_2+ \vartheta (\kappa _{2})\:q+
  \mathcal{O}(q^2) \,, \\
 \hat{\nu}(q) &=& -\vartheta (\kappa _{2})\frac{\tilde{g}(\mu_2)}
 {2 |\check{s}_{\beta}(\kappa_2)\!-\!\phi_a^0(\mu_2)|^2}\:
 q^2 +\mathcal{O}(q^3)\,.
 \end{eqnarray*}
\end{theorem}
\begin{proof} As in Theorem~\ref{resonth} we
rely on the implicit function theorem, but $\tilde\eta$ is now
jointly analytic, so is $z_2$. Since $\check{s}'_{\beta}
(\kappa_2)+2aK_0'(2a\kappa_2)>0$ the leading term of
$\hat{\nu}(q)$ is negative.
\end{proof}

\medskip

\begin{remark} \rm{The solution described in the theorem is not
unique, another one comes from the symmmetric eigenfunction of the
correponding Hamiltonian. This can be either a perturbed
eigenvalue if $\mu_1$ is isolated, or another resonance if $\mu_1$
is also embedded; in the threshold case, $\mu_1=-\frac{1}{4}
\alpha^2$, the behaviour depends on the sign of $q$.}
\end{remark}


\subsubsection{Symmetry broken by distance from the line}

Assume now on the contrary that the coupling strengths are the
same, $\beta \equiv \beta _{1}=\beta _{2}$, while one of the point
is shifted in the perpendicular direction, $x_{1}=(0,a)$ and
$x_{2}=(0,-a-\delta )$, where $\delta \in \R$. Now the equation
determining the resolvent pole acquires the form
 \begin{eqnarray*}
 \breve{\eta }(\delta ,z)&\!:=\!& s_{\beta }(z)^2-K_{0}((2a+\delta )
 \sqrt{-z})^2 - s_{\beta }(z)(\phi ^{l(z)}_{a+\delta }(z)
 +\phi ^{l(z)}_{a}(z)) \\ && -2K_{0}((2a+\delta )
 \sqrt{-z})\phi^{l(z)}_{a+\delta /2}(z)=0\,.
 \end{eqnarray*}
We keep notation $\kappa _{2}=\sqrt{-\mu_2}$ and put also
$$
f(\delta ,\kappa )=\breve{\eta }(\delta ,\sqrt{-\kappa ^{2}}).
$$
\begin{theorem}
Assume $0>\mu _{2}>-\frac{1}{4}\alpha^{2}$. For all nonzero and
sufficiently small $\delta $ the function $\breve{\eta }(\delta
,z)$ has a zero at a point $z(\delta )\in \Omega _{-}$ with the
real and imaginary part $z(\delta )=\upsilon (\delta )+ i\iota
(\delta )$ admitting the asymptotics
\begin{equation}\label{brosym}
\upsilon (\delta )= \mu_{2}-2 \kappa _{2}\kappa '_{2}\delta +
\mathcal{O} (\delta ^{2})\,,\quad  \iota (\delta )=-\kappa
_{2}\kappa ''_{2}\delta ^{2}+\mathcal{O} (\delta ^{3})\,,
\end{equation}
where
$$
\kappa'_{2}=-\frac{2aK'_{0}(2a\kappa _{2})}{\check{s}'_{\beta
}(\kappa_{2} ) +2aK'_{0}(2a\kappa _{2})}\,, \quad  \kappa
''_{2}=\frac{2f_{,\kappa  \delta }f_{,\delta }+f_{,\kappa  \kappa
}\kappa '_{2}-f_{,\delta \delta }f_{,\kappa }}{f_{,\kappa  }^{2}}
$$
and $f_{,i},\, f_{,ij}$ are appropriated derivatives at the point
$\{\delta ,\kappa\}= \{0,\kappa _{2}$\}. Moreover, we have $\iota
(\delta )<0$.
\end{theorem}
\begin{proof}
Similarly to Theorem~\ref{resonth} the argument is straightforward
being based on the implicit function theorem, hence we restrict
ourselves to commenting on the inequality $\iota (\delta )<0$. Let
$z(\delta )\in (-\frac{1}{4}\alpha ^2, 0)$. Without loosing
generality we can assume $\delta>0$ because the leading term of
$\iota (\delta )$ is quadratic in $\delta $, then $\kappa
_{2}(\delta )=\sqrt{-z(\delta )}=\kappa _{2}+\kappa '_{2}
\delta+\mathcal{O}(\delta ^2)>\kappa _{2}$. It is easy to see that
the first and the second component of $\breve{\eta }(\delta ,
z(\delta ))$ are real if the number $z(\delta )$ is real;
furthermore, using the explicit form for $\phi ^{l(z)}_{a}$ and
properties of the exponential function one can check that
$$
\mathrm{Im}\,f(\delta , \kappa_{2}(\delta ))<
-2\,\mathrm{Im}\,(g_{\alpha ,a+ \delta /2}(z(\delta
))(\check{s}_{\beta }(\kappa _{2}(\delta ))+K_{0} ((2a+\delta
)\kappa _{2}(\delta )))\,.
$$
Since we have $\check{s}_{\beta } (\kappa _{2}(\delta ))
+K_{0}((2a+\delta )\kappa _{2}(\delta ))>0$ the imaginary part of
$f(\delta ,\kappa _{2}(\delta ))$ is strictly negative.
Consequently, $z(\delta )$ can not be a real number, and the
possibility $\mathrm{Im }\,z(\delta )>0$ is excluded by general
spectral properties of self-adjoint operators.
\end{proof}


\subsection{Scattering } \label{scat}

While resonances in the analytically continued resolvent typically
coincide with poles of the continued scattering matrix, this
property does not hold automatically and has to be checked for
each particular system separately. Our next goal is to illustrate
it in the present setting, again in the simplest case with a
single point interaction localized at the point $y$. To this aim
we have thus to construct the $S$ matrix for the pair $(H_{\alpha
,\beta },\tilde{H}_{\alpha })$. Since the operator $H_{\alpha
,\beta }$ represents a rank-one perturbation of $\tilde{H}_{\alpha
}$, the existence and completeness of the corresponding wave
operators follows immediately from the Kuroda-Birman theorem.
Consequently, the $S$ matrix is unitary; our aim is to find the
on-shell S-matrix in the interval $(-\frac{1}{4}\alpha^2,0)$, i.e.
the corresponding transmission and reflection amplitudes.

\subsubsection{The on-shell S matrix}\label{Smatri}

Using the notation introduced above and Proposition~\ref{resoprop}
we can write the resolvent for $\mathrm{Im\,}z>0$ as
\begin{equation}\label{resapp}
 R_{\alpha ,\beta }(z)=R_{\alpha }(z) +\eta_a(z)^{-1}(\cdot\,,
 v_{z})v_{z},
\end{equation}
where the rank-one part in the last term is given by
$v_{z}:=R_{\alpha;L1}(z)$. We set $z=\lambda+i\varepsilon$ and
apply the operator $R_{\alpha ,\beta }(\lambda +i\varepsilon)$ to
$$\omega_{\lambda +i\varepsilon}(x):= \mathrm{e}^{i(\lambda
+i\varepsilon +\alpha^2/4)^{1/2}x_1}\, \mathrm{e}^{-\alpha
|x_{2}|/2}\,, $$
then we take the limit $\varepsilon\to 0+$ in the sense of
distributions. A straightforward if tedious calculation shows that
$H_{\alpha,\beta}$ has a generalized eigenfunction which for large
$|x_1|$ behaves as
\begin{eqnarray}\label{psibeh}
 \psi_\lambda(x) &\!\approx\!& \mathrm{e}^{i(\lambda+\alpha ^2/4)^{1/2}x_{1}}
 \, \mathrm{e}^{-\alpha|x_{2}|/2} \\ &&  +\frac{i}{8}\, \alpha
 \eta_a(\lambda)^{-1}\, \frac{\mathrm{e}^{-\alpha a}}{(\lambda
 +\frac{1}{4}\alpha^2)^{1/2}}\: \mathrm{e}^{i(\lambda
 +\alpha^2/4)^{1/2}|x_{1}| }\, \mathrm{e}^{-\alpha|x_{2}|/2}
 \nonumber
\end{eqnarray}
for each $\lambda\in(-\frac{1}{4}\alpha^2,0)$. To be more specific
about derivation of the above formula, one has to use again
(\ref{resoprop}) and to rely on considerations analogous to those
in the proof of Lemma~\ref{anacon} to arrive at
\begin{equation}\label{vlambd}
v_{\lambda }=\lim_{\varepsilon \to 0}v_{\lambda
+i\varepsilon}=R_{L1}(\lambda)+S(\lambda )\,,
\end{equation}
where
$$
S(\lambda)=I_{\lambda
}(x_{1},x_{2})+\frac{i}{8}\alpha\frac{\mathrm{e}^{-\alpha
(a+|x_{2}|)/2}}{(\lambda
 +\frac{1}{4}\alpha^2)^{1/2}}\: \mathrm{e}^{i(\lambda
 +\alpha^2/4)^{1/2}|x_{1}| }
$$
and
$$
I_{\lambda}(x_{1},x_{2}):=\mathcal{P}\int_0^{\infty}\frac{\mu^{0}
(\lambda,t)}{t-\lambda -\frac{1}{4}\alpha^2}\:\e
^{-|x_{2}|(t-\lambda)^{1/2}}\, \mathrm{e}^{it^{1/2}x_{1}}\,
\mathrm{d}t\,;
$$
here $\mu^{0} (\lambda,t)$ is defined in Sec.~\ref{resec1}.
Furthermore, note that the first component of (\ref{vlambd}) as
well as $I_{\lambda}(x_{1},x_{2})$ vanish for $|x_{1}|\to \infty$,
and at the same time
$$
\lim_{\varepsilon \to 0}(\omega _{\lambda
+i\varepsilon},v_{\lambda +i\varepsilon})=\e^{-\alpha a/2}.
$$
In view of the results of Sec.~\ref{resec1} and (\ref{resapp})
this yields formula (\ref{psibeh}) which, in turn, gives the
sought quantities (see also Appendix~B).
\begin{proposition} \label{scatres}
The reflection and transmission amplitudes are given by
 $$ \mathcal{R}(\lambda)= \mathcal{T}(\lambda)-1= \frac{i}{8}\,
 \alpha \eta_a(\lambda)^{-1}\, \frac{\mathrm{e}^{-\alpha a}}
 {(\lambda +\frac{1}{4}\alpha^2)^{1/2}}\,; $$
they have the same pole in the analytical continuation to the
region $\Omega_-$ as the continued resolvent.
\end{proposition}

\subsection{Unstable state decay}

It is also useful to look at the resonance problem from the
complementary point of view and to investigate the decay of an
unstable state associated with the resonance. Let us consider
again the simplest case $n=1$. The previous results tell us that
if the ``unperturbed'' eigenvalue $\epsilon_\beta $ of $H_{\beta}$
is embedded in $(-\frac{1}{4}\alpha^2, 0)$ and $a$ is large enough
then the corresponding resonance state has a long halflife. In
analogy with the Friedrichs model \cite{De} one might expect that
in the weak-coupling case, which corresponds to a large distance
$a$ here, the resonance state would be similar up to normalization
to the eigenvector $\xi_0 := K_{0}(\sqrt{ -\epsilon_\beta
}\,\cdot)$ of $H_\beta$ corresponding to $\epsilon_\beta$, with
the decay law being dominated by the exponential term.

However, the present model is different in one important aspect.
In a typical decay problem the decaying state belongs to the
absolutely continuous subspace of the Hamiltonian and thus the
decay law tends to zero as $t\to\infty$ by Riemann-Lebesgue lemma
\cite{Ex1}. Here we know from Sec.~\ref{disspe} that
$H_{\alpha,\beta}$ has always an isolated eigenvalue, and it is
easy to see that the latter is \emph{not} orthogonal to
$\psi_{\alpha,\beta,a}$ for any $a$; it is sufficient to realize
that both functions are positive, up to a possible phase factor.
Consequently, the decay law $|(\xi_0, U(t)\xi_0)|^2\|\xi_0\|^{-2}$
has always a nonzero limit as \mbox{$t\to\infty$} which is equal
to the squared norm of the projection of $\xi_0\|\xi_0\|^{-1}$ on
the eigensubspace given by $\psi_{\alpha,\beta,a}$. On the other
hand, this fact does not exclude that the decay \emph{is}
dominated by the natural exponential term as $a\to\infty$; it may
happen that the nonzero limit, which certainly depends on $a$, is
hidden in the non-exponential error term. This question requires a
longer discussion and we postpone it to a subsequent publication.


\section{Three dimensions: a plane and points} \label{plane+point}
\setcounter{equation}{0}

In analogy with the two-dimensional case investigated in the
previous sections we are going to discuss now briefly generalized
Schr\"{o}dinger operators in $L^2(\R^3)$ corresponding to the
formal expression
\begin{equation} 
-\Delta -\alpha \delta (x-\Lambda)+\sum _{i=1}^{n}\tilde\beta
_{i}\delta(x-y^{(i)})\,,
\end{equation}
where $\alpha >0$, $\beta _{i}\in \R $ and $ \Lambda
:=\{(\underline{x}_{1},0);\,\underline{x}_{1}\in\R^2 \}$ is a
plane, with $y^{(i)}\in \R^3 \setminus \Lambda $; for the point
set we will keep the same notation, $\Pi :=\{y^{(i)}\}_{i=1}^n$.

\subsection{Definition of Hamiltonian}

To write down appropriate boundary conditions let us consider
functions $f\in W^{2,2}_{\mathrm{loc}}(\R^{3}\setminus (\Lambda
\cup \Pi )) \cap L^{2}(\R^{3})$ which are continuous at $\Lambda
$. For any such function we put $f\upharpoonright
_{\mathcal{C}_{\rho ,i}}$ as its restriction to the points $x\in
\mathcal{C}_{\rho ,i}\equiv \mathcal{C}_{\rho}(y_{i}):=\{q\in
\R^3:|q-y^{(i)}|=\rho \}$. In analogy with the two-dimensional
case we set
$$
\Xi_{i}(f):=\lim _{\rho \rightarrow 0}\frac {1}{\rho }f
\upharpoonright _{\mathcal{C}_{\rho ,i}}\,,\quad
\Omega_{i}(f):=\lim _{\rho \rightarrow 0}[f\upharpoonright
_{\mathcal{C}_{\rho ,i}} -\Xi_{i}(f)\rho ]
$$
for $i=1,...,n$, and
$$
\Xi_{\Lambda }(f)(x_{1}):= \partial_{x_{2}} f(x_{1},0+)-
\partial_{x_{2}}f(x_{1},0-)\,, \quad \Omega_{\Lambda }(f)(x_{1}):=
f(x_{1},0)\,,
$$
and we assume that the above limits are finite and satisfy the
relations
\begin{equation}  \label{boucon2}
\Xi _{i}(f)=4\pi \beta _{i}\Omega  _{i}(f)\: , \quad \Xi _{\Lambda
}(f)(x_{1})=-\alpha \Omega  _{\Lambda}(f)(x_{1})\,.
\end{equation}
Then we define $H_{\alpha, \beta}$ as the Laplace operator with
the boundary conditions given now by (\ref {boucon2}); it is
straightforward to check that it is self-adjoint on its natural
domain.

\subsection{Resolvent of $H_{\alpha, \beta}$}

In the three-dimensional case the free resolvent $R(z)$ with $z\in
\rho(-\Delta )$ is an integral operator in $L^{2}(\R^{3})$ having
the kernel
\begin{equation} \label{kerneG3}
G_{z}(x,x')=\frac{1}{(2\pi)^{3}}\int _{\R^{3}}\frac{\mathrm{e}^
{ip(x-x')}}{p^{2}-z}\,\mathrm{d}p =
\frac{\mathrm{e}^{i\sqrt{z}}|x-x'|}{4\pi|x-x'|}\,.
\end{equation}
Now we introduce the auxiliary Hilbert spaces
$\mathcal{H}_{0}\equiv L^{2}(\R^{2}), \mathcal{H}_{1}\equiv \C^n$
and abbreviate $L^2\equiv L^2(\R^3)$, $W^{2,2}\equiv
W^{2,2}(\R^3)$. By means of the trace maps $\tau _0:W^{2,2}\to
\mathcal{H}_0$ and $\tau _1 : W^{2,2}\to \mathcal{H}_1$ acting as
 $$
 \tau _0  f:=f\!\upharpoonright_{\,\Lambda }\,, \quad \tau_1
 f:=f\!\upharpoonright_{\,\Pi}=(f\!\upharpoonright _{\,\{y^{(1)}\}},\dots,
 f\!\upharpoonright _{\,\{y^{(n)}\}})\,,
 $$
we define in analogy with (\ref{embedd}) the embeddings
$\mathbf{R}_{iL}(z)$, $ \mathbf{R}_{Li}(z)$, and $
\mathbf{R}_{ji}$. The operator-valued matrix $\Gamma (z)$ now
takes the form
$$
\Gamma (z)=[\Gamma
_{ij}(z)]:\mathcal{H}_{0}\oplus\mathcal{H}_{1}\rightarrow
\mathcal{H}_{0}\oplus\mathcal{H}_{1}\,,
$$
where $\Gamma _{ij}(z):\mathcal{H}_{i}\rightarrow \mathcal{H}_{j}$
are the operators given by
 \begin{eqnarray*} \label{forg22plane}
 \Gamma _{ij}(z)g &=& -\mathbf{R}_{ij}(z)g \qquad
 \mathrm{for}\quad i\neq j \quad \mathrm{and }\quad g\in
 \mathcal{H}_{j}\,, \\ \Gamma_{00}(z)f &=& \left[\,\alpha^{-1}
 \!-\mathbf{R}_{00}(z)\right] f \qquad \mathrm{if} \quad
 f\in \mathcal{H}_0\,, \\ \Gamma _{11}(z)\varphi &=&
 \left[ \left(\beta _{l}+\frac{i\sqrt{z}}{4\pi}\right) \delta_{kl}
 - G_{z}(y^{(k)},y^{(l)}) (1\!-\!
 \delta_{kl}) \right]_{k,l=1}^{n} \varphi \quad \mathrm{for}
 \quad \varphi \in \mathcal{H}_1\,.
 \end{eqnarray*}
To describe the inverse of $\Gamma (z)$ we introduce the reduced
determinant $D(z)\equiv D_{11}(z):\mathcal{H}_{1}\rightarrow
\mathcal{H}_{1}$ given again by $D(z)=\Gamma _{11}(z)-\Gamma
_{10}(z)\Gamma _{00}(z)^{-1}\Gamma _{01}(z)$ for $z$ belonging to
the resolvent set of $H_{\alpha ,\beta }$. The inverse of $\Gamma
(z)$ is given by $
[\Gamma(z)]^{-1}:\mathcal{H}_{0}\oplus\mathcal{H}%
_{1}\rightarrow \mathcal{H}_{0}\oplus\mathcal{H}_{1}$ defined as
in (\ref{invgam}). Calculations similar to those of
Theorem~\ref{resoth} yield the resolvent formula for $z\in \rho
(H_{\alpha ,\beta })$ and $\mathrm{Im \,}z>0$ in the form
\begin{equation} 
R_{\alpha ,\beta }(z)\equiv (H_{\alpha ,\beta
}-z)^{-1}=R(z)+\sum_{i,j=0}^{1} \mathbf{R}
_{Li}(z)[\Gamma(z)]_{ij}^{-1}\mathbf{R}_{jL}(z)\,.
\end{equation}

\subsection{Spectrum of $H_{\alpha ,\beta }$}

Since the point interactions give rise to an explicit finite-rank
perturbation to the resolvent, we find easily the absolutely
continuous spectrum,
$$
\sigma_\mathrm{ess}(H_{\alpha ,\beta })
=\sigma_\mathrm{ac}(H_{\alpha ,\beta
})=[-\frac{1}{4}\alpha^{2},\infty )\,.
$$
As for the discrete spectrum we start again with the simplest case
of a single point perturbation located at a distance $a$ from
$\Lambda $; the coupling constant of this interaction is $\beta
\in \R$. As we have said in the introduction we will concentrate
only on the differences coming from the fact that the relative
dimension of the two components of the interaction support is now
two.

Let us denote by $H_{\beta }\equiv H_{0,\beta }$ the Laplace
operator in $L^ {2}$ with the perturbation supported at $y$ only.
It is well known \cite{AGHH} that if $\beta <0$ then the
Hamiltonian $H_{\beta }$ has a single eigenvalue given by
$$
\tilde{\epsilon }_{\beta}=-(4\pi \beta )^{2}.
$$
In turn, if $\beta \geq 0$ the spectrum of $H_{\beta}$ has no
isolated point. However, as we will see below,  the operator
$H_{\alpha ,\beta }$ with $\alpha >0$ has an eigenvalue even in
the latter case. To derive spectral properties of $H_{\alpha
,\beta }$ we have to find solutions of the equation
$\check{D}(\kappa )=0$ for $\kappa \in (\frac{1}{2}\alpha
,\infty)$, where the operator $\check{D}(\kappa )$ now acts as the
multiplication by the following function,
$$
\check{d}_{a}(\kappa):= \beta +\frac{ \kappa }{4\pi
}-\check{\phi}_{a}(\kappa )
$$
with
$$
\check{\phi}_{a}(\kappa ):=\frac{\alpha }{\pi }\int_{0}^{\infty
}\frac{ \mathrm{e}^{-2(p^{2}+\kappa ^{2})^{1/2} a
}}{(2(p^{2}+\kappa ^{2})^{1/2}-\alpha )(p^{2}+\kappa
^{2})^{1/2}}\:p\,\mathrm{d}p\,.
$$
Since we want to investigate simultaneously the asymptotics of the
eigenvalue for large and small $a$ it is convenient to put
$H_{\alpha , \beta , a}=H_{\alpha , \beta }$. We have

\begin{theorem} \label{isosp3}
For any $\alpha >0$ and $\beta \in \R$ the operator $H_{\alpha
,\beta ,a }$ has exactly one isolated eigenvalue
$-\kappa_{a}^{2}<-\frac{1}{4}\alpha ^{2}$. Moreover, if $\beta >0$
or $\tilde{\epsilon}_{\beta}\in [-\frac{1}{4}\alpha ^{2},\infty )$
then
\begin{equation}\label{conve1}
-\lim _{a\to \infty }\kappa _{a}^{2}=\tilde{\epsilon}_{\beta}\,,
\end{equation}
otherwise we have
\begin{equation}\label{conve2}
-\lim _{a\to \infty }\kappa _{a}^{2}=-\frac{1}{4}\alpha ^{2}\,.
\end{equation}
In distinction to the two-dimensional situation we have now
\begin{equation}\label{conve4}
-\lim _{a\to 0}\kappa _{a}^{2}= -\infty \,.
\end{equation}
\end{theorem}
\begin{proof}
The equations (\ref{conve1}) and (\ref{conve2}) can be obtained by
mimicking the arguments employed in proofs of Thms~\ref{isospe}
and \ref{asymev}. Using the explicit form for $\check{\phi}_{a}$
one can establish the existence of a positive $C$ such that
$Ca^{-1}<\check{\phi }_{a}(\kappa )$. It follows that $\lim _{a\to
0}\check{\phi }_{a}(\kappa )=\infty$ which, in turn, implies
(\ref{conve4}).
\end{proof}
\begin{remark}
\rm{ In the three-dimensional case one may say that the behaviour
of the eigenvalue for large $a$ depends not only on the relation
between $-\alpha ^2 / 4$ and  $\tilde\epsilon_{\beta }$; in the
limit it is absorbed in the threshold also in the case when
$\beta\ge 0$ and the discrete spectrum of $H_\beta$ is empty.}
\end{remark}

Proceeding similarly as in the proof of Theorem~\ref{evn-ca}
arrive at
\begin{theorem} 
Let $\beta=(\beta  _{1},...,\beta  _{n})$, where $\beta _{i}\in
\R$ and $\alpha >0$. Operator $H_{\alpha ,\beta }$ has at least
one isolated eigenvalue and at most $n$. If all the numbers
$-\beta _{i}$ are sufficiently large then $H_{\alpha ,\beta }$ has
exactly $n$ eigenvalues.
\end{theorem}

\subsection{Resonances}

To recover the resonances for the model in question we can proceed
similarly as in Sec.~\ref{resec1}. Assume that $\beta <0$ and
$\tilde{\epsilon}_{\beta }>-\alpha ^2 /4$. In analogy with
Lemma~\ref{anacon} we state that the resolvent of $H_{\alpha ,
\beta }$ has a second-sheet continuation through the interval
$(-\frac{1}{4}\alpha^2, 0)$. Let us put $\tilde{\varsigma}_{\beta
}:=\sqrt{- \tilde{\epsilon} _{\beta }}=4\pi \beta $.

\begin{theorem} 
Assume $\tilde{\epsilon}_{\beta }> -\frac{1}{4}\alpha^2$. For any
$a$ sufficiently large the resolvent $R_{\alpha ,\beta }$ has the
second sheet pole at a point $z(a)$ with the real and imaginary
part, $z(a)=\mu(a)+ i\nu(a) ,\; \nu (a)<0,$ which in the limit
$a\to\infty$ behave in the following way,
 \begin{equation}\label{assyRI2}
 \mu(a
)=\tilde{\epsilon}_{\beta}+\mathcal{O}(\mathrm{e}^{- a \tilde
{\varsigma}_{\beta }} )\,, \quad \nu(a)
 =\mathcal{O}(\mathrm{e}^{- a \tilde{\varsigma}_{\beta }})\,.
 \end{equation}
 \end{theorem}

\begin{remark}
\rm{The resonance pole exists even if the distance is not large.
In contrast to the two dimensional case, however, the imaginary
part of the pole position $\nu (a)$ diverges to $-\infty $ as
$a\to 0$.}
\end{remark}


\setcounter{section}{1} \setcounter{equation}{0}
\renewcommand{\theequation}{\Alph{section}.\arabic{equation}}
\renewcommand{\theclaim}{\Alph{section}.\arabic{equation}}
\section*{Appendix A: Proof of Lemma~\ref{anacon}}

In view of the edge-of-the-wedge theorem, our aim is to show that
\begin{equation}
\lim_{\varepsilon \rightarrow 0^{+}}\phi _{a}^{\pm }(\lambda \pm
i\varepsilon )=\phi _{a}^{0}(\lambda )\quad \mathrm{for} \quad
-\frac{1}{4}\alpha^{2} < \lambda <0\,. \label{ep+-=0}
\end{equation}
Given $\varepsilon >0$ we put $z_{\lambda }^{\pm }(\varepsilon ):=
\lambda \pm i\varepsilon$. Let $\delta (\cdot)$ be function of the
parameter $\varepsilon $ such that $0<\delta(\varepsilon
)<\varepsilon $. We use them to define a family of the sets
$C_{i}^{\pm}(\varepsilon)$ in the complex plane, each of which may
be regarded as a graph of a curve,
\begin{eqnarray*}
C_{1}(\varepsilon ) &\!\equiv\!& C_{1}^{\pm}(\varepsilon )
:=\{w=x\,:\: x\in [\delta (\varepsilon ),\varepsilon ^{-1}]\}\,,
\\ C_{2}^{\pm}(\varepsilon ) &\!:=\!& \{w=x\pm i\varepsilon
\,:\: x\in [0,x_{2}]\cup[x_{1},\varepsilon ^{-1}] \}
\end{eqnarray*}
with
$$ x_{k}\equiv x_{k}(\varepsilon ):=\lambda +
\frac{1}{4}\alpha^2 +(-1)^{k+1}\delta (\varepsilon )\,,\quad
k=1,2\,\:;
$$
furthermore,
 \begin{eqnarray*}
 C_{3}^{\pm}(\varepsilon )&\!:=\!& \{w
=z^{\pm}(\varepsilon)+\frac{1}{4}\alpha^2 +\delta
(\varepsilon)\e^{i\theta }:\: \theta \in\mp[0
,\pi ]\}\,, \\
C_{4}^{\pm}(\varepsilon )&\!:=\!& \{w=\varepsilon^{-1}\pm iy:\:
y\in [0,\varepsilon]\} \cup\{w=\pm iy:\:\, y\in
[\delta(\varepsilon),\varepsilon ]\}\,, \\
C_{5}^{\pm}(\varepsilon )&\!:=\!& \{w= \delta(\varepsilon)
\e^{i\theta }:\: \pm\theta \in [0,\frac{1}{4}\pi]\}\,.
 \end{eqnarray*}
It is easy to see that from the definitions of $C_{l}^{\pm
}(\varepsilon )$ that each of their unions,
 $$
C^{\pm }(\varepsilon ):= \sum_{l=1}^{5}C_{l}^{\pm }
(\varepsilon)\,, $$
is a graph of a closed curve in the closed upper and lower complex
halfplane, respectively, and that the regions encircled by these
loops do not contain singularities of the functions $w\mapsto \mu
(z_{\lambda }^{\pm }(\varepsilon ),w)(w-z_{\lambda }^{\pm
}(\varepsilon )-\frac{1}{4}\alpha^{2} )^{-1}$; thus by the basic
theorem about analytic functions we have
\begin{equation}
\int_{C^{\pm }(\varepsilon )}\frac{\mu (z_{\lambda }^{\pm
}(\varepsilon ),w) }{(w-z_{\lambda }^{\pm }(\varepsilon
)-\frac{1}{4}\alpha^{2})}\,\D w=0\,. \label{analymu}
\end{equation}
This will be our starting point to check the relation
(\ref{ep+-=0}):

\emph{1st step}: Since by assumption $\delta(\varepsilon)\to 0$ as
$\varepsilon \rightarrow 0^{+}$ so $C_{1}(\varepsilon )$
approaches the positive real halfline, the limits we want to find
are equal
$$ \lim_{\varepsilon \rightarrow 0^{+}}\phi _{a}^{+}(z_{\lambda
}^{+ }(\varepsilon ))=\lim_{\varepsilon \rightarrow
0^{+}}\int_{C_{1}(\varepsilon )}\frac{\mu (z_{\lambda }^{+
}(\varepsilon ),w)}{ w-z_{\lambda }^{+ }(\varepsilon
)-\frac{1}{4}\alpha^{2}}\,\D w $$
and
$$ \lim_{\varepsilon \rightarrow 0^{+}}\phi _{a}^{-}(z_{\lambda
}^{- }(\varepsilon ))=-\lim_{\varepsilon \rightarrow
0^{+}}\int_{C_{1}(\varepsilon )}\frac{\mu (z_{\lambda }^{-
}(\varepsilon ),w)}{ w-z_{\lambda }^{-}(\varepsilon
)-\frac{1}{4}\alpha^{2}}\,\D w +g^{-}_{\alpha , a}(z^{-}_{\lambda
}(\epsilon))\,. $$

\emph{2nd step}: Consider next the integration over $w^{\pm }=t\pm
i\eta (\varepsilon )\in C_{2}^{\pm }(\varepsilon ).$ Using the
following obvious convergence relations,
\begin{eqnarray*}
(z_{\lambda }^{\pm }(\varepsilon )-w^{\pm })^{1/2} &\to
&i(t-\lambda )^{1/2}\quad \mathrm{as}\quad \varepsilon \rightarrow
0\,, \\ \sqrt{w^{\pm }} &\to &\pm \sqrt{t}\quad\;
\mathrm{as}\quad\; \varepsilon \rightarrow 0\,,
\end{eqnarray*}
we find
\begin{equation}
\lim_{\varepsilon \rightarrow 0^{+}}\int_{C_{2}^{\pm }(\varepsilon
)}\frac{ \mu (z_{\lambda }^{\pm }(\varepsilon ),w^{\pm })}{w^{\pm
}-z_{\lambda }^{\pm }(\varepsilon )-\frac{\alpha ^{2}}{4}}\,\D
w^{\pm }=\mp\mathcal{P} \int_{0}^{\infty }\frac{\mu ^{0}(\lambda ,
t)}{t-\lambda -\frac{1}{4}\alpha^{2}}\,\D t\,.
\end{equation}

\emph{3rd step}: In the integration over the circular segments
around the poles away of the origin, $w^{\pm }\in C_{3}^{\pm
}(\varepsilon )$, we employ the convergence
\begin{eqnarray*}
(z_{\lambda }^{\pm }(\varepsilon )-w^{\pm })^{1/2} &\rightarrow
&\frac{i}{2}\alpha \quad\; \mathrm{as}\quad\; \varepsilon \to 0
\\ \sqrt{w^{\pm }} &\rightarrow &\pm \sqrt{\lambda
+\frac{1}{4}\alpha^{2}} \quad \mathrm{as}\quad
\varepsilon \rightarrow 0
\end{eqnarray*}
which yields
\begin{equation}
\mu (z_{\lambda }^{\pm }(\varepsilon ),w^{\pm })\rightarrow \pm\:
\frac{ g_{\alpha ,a}(\lambda )}{\pi i}\quad \mathrm{as}\quad
\varepsilon \rightarrow 0\,.  \label{convmu}
\end{equation}
To proceed further we use the following identities
\begin{eqnarray*}
\lefteqn{ \int_{C_{3}^{\pm }(\varepsilon )}\frac{\mu (z_{\lambda
}^{\pm }(\varepsilon ),w^{\pm })}{w^{\pm }-z_{\lambda }^{\pm
}(\varepsilon )-\frac{1}{4}\alpha^{2}}\,\D w^{\pm } = \pm\,
\frac{g_{\alpha ,a}(\lambda )}{\pi i}\int_{ C_{3}^{\pm
}(\varepsilon )}\frac{1}{w^{\pm }-z_{\lambda }^{\pm }(\varepsilon
)-\frac{1}{4}\alpha^{2}}\,\D w^{\pm } } \\ && \phantom{AAAAAA}
+\int_{C_{3}^{\pm }(\varepsilon )}\frac{\mu (z_{\lambda }^{\pm
}(\varepsilon ),w^{\pm })\mp g_{\alpha ,a}(\lambda )(\pi
i)^{-1}}{w^{\pm }-z_{\lambda }^{\pm }(\varepsilon
)-\frac{1}{4}\alpha^{2}}\,\D w^{\pm }\,. \phantom{AAAAAAA}
\end{eqnarray*}
Since $\lim_{\varepsilon \rightarrow 0}\int_{C_{3}^{\pm
}(\varepsilon )}\frac{1}{w^{\pm }-z_{\lambda }^{\pm }(\varepsilon
)-\frac{1}{4}\alpha^{2}} \,\D w^{\pm }=\mp\pi i$, the limit as
$\varepsilon \to 0^+$ of the first component in the above relation
equals $\mp\, g_{\alpha ,a}(\lambda )$. Moreover, in view of the
convergence (\ref{convmu}) and the fact that the functions
involved are continuous at the segment in question we can find a
function $\varepsilon \mapsto \zeta (\varepsilon )$ such that
$\zeta (\varepsilon )\rightarrow 0$ as $\varepsilon \rightarrow 0$
and $\left| \mu (z_{\lambda }^{\pm }(\varepsilon ),w^{\pm })\mp
g_{\alpha ,a}(\lambda ) (\pi i)^{-1}\right| <\zeta (\varepsilon )$
for $w^{\pm }\in C_{3}^{\pm }(\varepsilon )$. Then
$$ \int_{C_{3}^{\pm }(\varepsilon )}\left| \frac{\mu (z_{\lambda
}^{\pm }(\varepsilon ),w^{\pm })\mp g_{\alpha ,a}(\lambda
)(4i)^{-1}} {w^{\pm }-z_{\lambda }^{\pm }(\varepsilon
)-\frac{1}{4}\alpha^{2}}\right| \,\D w^{\pm }<\pi \zeta
(\varepsilon )\,, $$
i.e. the second integral in the above identity vanishes as
$\varepsilon \rightarrow 0.$ Summarizing the argument we get
$$ \lim_{\varepsilon \rightarrow 0^{+}}\int_{C_{3}^{\pm
}(\varepsilon )} \frac{\mu (z_{\lambda }^{\pm }(\varepsilon
),w^{\pm })}{w^{\pm }-z_{\lambda }^{\pm }(\varepsilon
)-\frac{1}{4}\alpha^{2}}\,\D w^{\pm }=- g_{\alpha ,a}(\lambda )\,.
$$

\emph{4th and 5th step}: Next we note that the limit
$|w^{\pm}|\frac{ \mu (z_{\lambda }^{\pm }(\varepsilon ),w^{\pm
})}{w^{\pm }-z_{\lambda }^{\pm }(\varepsilon
)-\frac{1}{4}\alpha^{2}}$ as $\varepsilon \to 0 $ implies for the
integral over the ``vertical'' parts of the integration curve
$$ \lim_{\varepsilon \rightarrow 0^{+}}\int_{C_{4}^{\pm
}(\varepsilon )}\frac{ \mu (z_{\lambda }^{\pm }(\varepsilon
),w^{\pm })}{w^{\pm }-z_{\lambda }^{\pm }(\varepsilon
)-\frac{1}{4}\alpha^{2}}\,\D w^{\pm }=0\,. $$
Finally, it is also easy to see that the remaining integral over
$C_{5}^{\pm }(\varepsilon )$ vanishes in the limit $\varepsilon
\rightarrow 0$. Combining (\ref{analymu}) with the above results
we get
$$ \lim_{\varepsilon \rightarrow 0^{+}}\phi _{a}^{\pm }(z_{\lambda
}^{\pm }(\varepsilon ))=\phi _{a}^{0}(\lambda )\,, $$
so the function $\phi _{a}^{0}$ is continuous for $\lambda \in
(-\frac{1}{4}\alpha^{2},0)$ and the proof is complete.


\setcounter{section}{2} \setcounter{equation}{0}
\renewcommand{\theequation}{\Alph{section}.\arabic{equation}}
\renewcommand{\theclaim}{\Alph{section}.\arabic{equation}}
\section*{Appendix B: Lippmann--Schwinger equation }

Here we present another possible approach to the scattering
problem which we have discussed in Sec.~\ref{scat}.

\subsubsection{Additive representation of $H_{\alpha ,\beta }$}

It is also useful to write $H_{\alpha ,\beta }$ in an additive
form which would be remiscent of the usual potential interaction
-- cf.~\cite{KK, Ko, AP1}. To this aim, let us construct for the
operator $\tilde{H}_{\alpha }:D(\tilde{H}_{\alpha })\to L^2$ the
natural rigged Hilbert space, i.e. the triplet
$$ \mathcal{H}_{\alpha;-}\supset L^2 \supset
\mathcal{H}_{\alpha;+}\:, $$
where $\mathcal{H}_{\alpha;\pm}$ are the completion of
$D(\tilde{H}_{\alpha })$ in the norm
$$ \|f\|_{\pm}:= \|(\tilde{H}_{\alpha }\! -\!\lambda)^{\pm
1}f\|\,,\quad \mathrm{where }\quad \lambda<-\frac{1}{4}\alpha^2.
$$
Then we can define the extension of $\tilde{H}_{\alpha }$ to whole
$L^2$; this leads to the map $\mathbf{H}_{\alpha }:L^2 \to
\mathcal{H}_{\alpha;-}$ which expresses the canonical unitarity
between $L^2$ and $\mathcal{H}_{\alpha;-}$. Let $D(V_{\beta })$
denote the set of functions $f\in W^{2,2}_{\mathrm{loc}}(\R^{2}
\setminus (\Sigma \cup\Pi )) \cap L^{2}$ such that the limits $\Xi
_{\Sigma }(f)$, $\Omega _{\Sigma }(f)$ satisfy (\ref{boucon}) and
$\Xi _{i}(f)$, $\Omega _{i}(f)$ are finite.  Now we define the
operator $V_{\beta }:D(V_{\beta })\to \mathcal{H}_{\alpha ;-} $ by
$$ V_{\beta }\psi =\sum _{i=1}^{n}\psi ^{\beta _{i}}_{reg}\delta
(\cdot -y ^{(i)})\,,\;\; \mathrm{where} \quad \psi ^{\beta
_{i}}_{reg} := \left\{
\begin{array}{cc}
-(2\pi \beta _{i})^{-1}\Omega _{i}(\psi) & \mathrm{if} \quad \beta
\neq 0 \\ -\Xi _{i}(\psi) & \mathrm{if } \quad \beta =0
\end{array} \right.
$$
Let us note that since $\mathbf{R}_{\alpha ;L1}= \sum _{i=1}^ {n}
G^{(\alpha )}_{z}\ast \delta (\cdot -y^ {(i)})\in L^2$ the
operator $V_{\beta }$ is indeed well defined as a map acting to
$\mathcal{H}_{\alpha ;-}$. Now we can define the sought operator,
\begin{equation}\label{gensum}
\tilde{H}_{\alpha }\hat{+}V_{\beta }:D( \tilde{H}_{\alpha
}\hat{+}V_{\beta })\to L^2,\quad ( \tilde{H}_{\alpha
}\hat{+}V_{\beta })f=\mathbf{H}_{\alpha }f+V_{\beta }f\,,
\end{equation}
with the domain given by
$$ D( \tilde{H}_{\alpha }\hat{+}V_{\beta })=\{g\in D(V_{\beta }):
\mathbf{H}_{\alpha }g+V_{\beta }g\in L^2 \}\,. $$
With this notations we have the following result.
\begin{lemma}
$\:H_{\alpha ,\beta }=\tilde{H}_{\alpha }\hat{+}V_{\beta }\,$.
\end{lemma}
\begin{proof}
It is easy to see that $\mathbf{H}_{\alpha }g+V_{\beta }g\in L^2 $
if and only if $g\in D(H_{\alpha ,\beta })$ because only the
boundary conditions given by (\ref{boucon}) ensure the appropriate
compensation of $\delta (\cdot -y^ {(i)})$ induced by $V_{\beta}$
-- cf.~\cite{Ko}. At the same time, it is also easy to see that
$(\tilde{H}_{\alpha }\hat{+}V_{\beta })g(x) =\tilde{H}_{\alpha
}g(x)$ for $x\in \R^{2}\setminus \Pi $; this completes the proof.
\end{proof}


\subsubsection{Generalized Lippman--Schwinger equation}

In the same vein we want to find now an analog of the
Lippman--Schwinger equation -- cf.~\cite{ABK}. The additive
representation (\ref{gensum}) provides an inspiration: it is
reasonable to expect that the generalized eigenvectors $\psi
^{\pm}_\lambda $ of $H_{\alpha ,\beta }$ will satisfy
\begin{equation}\label{genLSc}
\psi ^{\pm}_\lambda =\omega _\lambda -R^{\pm }_\alpha (\lambda
)V_{\beta }\psi ^{\pm}_\lambda  \quad \mathrm{for }\quad \lambda
\in [-\frac{1}{4}\alpha ^{2}, \infty )\,,
\end{equation}
where $\omega _\lambda =\lim_{\varepsilon \to 0} \omega_{\lambda
+i\varepsilon }$ are the generalized eigenvectors of $H_{\alpha }$
introduced in Sec.~\ref{Smatri} and $R^{\pm }_\alpha (\lambda )$
are the limits $\lim _{\varepsilon \to 0^{+}}R_\alpha (\lambda \pm
i\varepsilon )$ in a suitable generalized sense. We have to
emphasize that the equation (\ref{genLSc}) has only a formal
meaning; our aim is now to replace it by a mathematically rigorous
object. For $z^{\pm}(\varepsilon)=\lambda \pm i\varepsilon $
define functions $\psi_{z^{\pm}(\varepsilon)}\in L^2$ by
\begin{equation} \label{geneig}
\psi _{z^{\pm}(\varepsilon)}:=(H_{\alpha
,\beta}-z^{\pm}(\varepsilon))^{-1}(\tilde{H}_{\alpha
}-z^{\pm}(\varepsilon))\omega _{z^{+}(\varepsilon)}\,,
\end{equation}
i.e. the limits $\psi _{\lambda }^{\pm}:=\lim_{\varepsilon \to
0}\psi _{z^{\pm}(\varepsilon)}$ in the distributional sense
constitute the generalized eigenvalues of $H_{\alpha ,\beta}$.
Furthermore, a direct calculation shows the following relation
\begin{equation}\label{LSauxi}
\psi_{z^{\pm}(\varepsilon)}:=\omega _{z^
{+}(\varepsilon)}-R_{\alpha } (z^{\pm}(\varepsilon))V_{\beta
}\psi_{z^{\pm}(\varepsilon)}.
\end{equation}
which after taking the distributional limit $\varepsilon \to 0$
gives the strict meaning to heuristic relation (\ref{genLSc}). Of
course, the limits $\psi^{\pm}_{\lambda }$ belong only locally to
$L^2$, however, they satisfy the same boundary conditions on
$\Sigma \cup \Pi$ as functions from $D(H_{\alpha ,\beta })$. This
allows us to construct the extension $\bar{V}_\beta $ of $V_\beta
$ to $\psi ^{\pm}_\lambda$ because the latter ``feels" only the
behaviour of functions on $\Pi$. With this notation the relation
(\ref{LSauxi}) after taking the limit $\varepsilon \to 0$ acquires
the following form,
\begin{equation}\label{LipSch}
\psi ^{\pm}_\lambda =\omega_\lambda -R^{\pm}_\alpha (\lambda)
\bar{V}_\beta \psi ^{\pm}_\lambda \,.
\end{equation}

\subsection*{Acknowledgments}

S.K. is grateful for the hospitality in the Department of
Theoretical Physics, NPI, Czech Academy of Sciences, where a part
of this work was done. The research has been partially supported
by GAAS under the contract A1048101 and Polish Ministry of
Scientific Research and Information Technology under the
(solicited) grant No PBZ-MIN-008/P03/2003.


\end{document}